%% file: crealnvp_db.tex
\documentclass[aip, reprint, floatfix]{revtex4-1}

\usepackage{mystyle}
\input{acronyms}

\begin{document}

%\preprint{}

% Extended Machine Learning Scattering Kernel for very low Earth orbit: From Thermal to Orbital Velocities
\title[Conditional Normalizing Flow Scattering Kernel]{Conditional Normalizing Flow for Gas-Surface Scattering from Thermal to Hypersonic Velocities} %Title of paper

\author{Miklas Schütte}
\email{schuettem@irs.uni-stuttgart.de}
\thanks{Corresponding author}
\affiliation{Institute of Space Systems, University of Stuttgart}%
\homepage{https://www.irs.uni-stuttgart.de/en/}

\author{Stephen Hocker}%
\affiliation{Institute for Functional Matter and Quantum Technologies, University of Stuttgart}
\homepage{https://www.fmq.uni-stuttgart.de/}

\author{Hansjörg Lipp}
\affiliation{Institute for Functional Matter and Quantum Technologies, University of Stuttgart}

\author{Johannes Roth}
\affiliation{Institute for Functional Matter and Quantum Technologies, University of Stuttgart}

\author{Stefanos Fasoulas}
\affiliation{Institute of Space Systems, University of Stuttgart}

\author{Marcel Pfeiffer}
\affiliation{Institute of Space Systems, University of Stuttgart}

% \date{\today}
\date{\today, Preprint submitted but not yet peer-reviewed.}

% ----------------------------------------------------------------------------------------------- %
% Abstract
% ----------------------------------------------------------------------------------------------- %
\begin{abstract}
    Accurate aerodynamic modeling of satellites in \gls{vleo} requires \gls{gsi} models that capture the full velocity spectrum from thermal to orbital speeds.
    Atmospheric particles initially strike spacecraft surfaces at hypersonic velocities of $6\,000\,$-$\,10\,000\,\text{m/s}$.
    Due to surface roughness and complex geometries, especially within \gls{abep} intake systems, multiple collisions occur, progressively reducing the particle velocities.
    A recent machine learning framework for deriving scattering kernels from \gls{md} simulations has shown promise, but remains limited to high-velocity single impacts and possibly violates fundamental equilibrium principles such as detailed balance.
    This work extends this machine learning based scattering kernel to cover the complete velocity range using conditional normalizing flows trained with physics-informed constraints, enabling accurate modeling of multi-bounce scenarios in realistic \gls{vleo} applications.
    We train a \gls{crealnvp} model on expanded molecular dynamics simulations covering velocities from thermal to hypersonic speeds, incorporating a detailed balance loss term.
    The resulting model demonstrates improved accuracy compared to previous approaches even in the original high-velocity regime, while successfully capturing thermal-velocity scattering.
    Quantitative assessment shows that thermalization is approximated within acceptable tolerances.
    This framework provides essential capabilities for accurate \gls{abep} intake optimization and \gls{vleo} mission planning while offering a general methodology applicable to broader rarefied gas dynamics problems requiring thermodynamic consistency.
\end{abstract}

\glsresetall % reset all glossary entries to ensure they are defined in the text


% ----------------------------------------------------------------------------------------------- %
% Title
% ----------------------------------------------------------------------------------------------- %
\maketitle %\maketitle must follow title, authors, abstract and \pacs

% ----------------------------------------------------------------------------------------------- %
% Introduction
% ----------------------------------------------------------------------------------------------- %
\section{Introduction} \label{sec:introduction}

Describing particle surface scattering is particularly important in the free molecular flow regime.
The \gls{vleo}, which is typically defined for orbit altitudes below $450\,\text{km}$, falls entirely within this regime~\cite{hao_very_2014,crisp_benefits_2020}.
At these altitudes, the aerodynamic forces acting on satellites arise from \glspl{gsi} between the residual atmosphere and the surface of the spacecraft.
These interactions generate aerodynamic drag, which limits the lifetime of satellites in \gls{vleo}.
At the same time, they also enable novel concepts such as \gls{abep}, where an intake collects incoming particles to be used as propellant, and aerodynamic attitude and orbit control.
Therefore, accurate modeling of \glspl{gsi} is essential for lifetime prediction, mission planning, and aerodynamic optimization of satellite designs in \gls{vleo}.
Addressing these challenges is a central objective of the collaborative research center ATLAS (Advancing Technologies for Low-Altitude Satellites), which aims to make sustainable operations in \gls{vleo} feasible~\cite{fasoulas_atlas_2025}.

Widely used scattering models in \gls{vleo} aerodynamic simulation tools, such as \gls{dsmc} or panel methods, are very simple.
The standard scattering model is the Maxwell model, which was introduced by James Clerk Maxwell in 1879~\cite{Maxwell_stresses_1879}.
It predefines the scattering to be specular or diffuse.
The decision of a specular or diffuse scattering is decided by an accommodation coefficient.
However, molecular beam experiments have shown that the scattering behavior is much more complex and cannot be represented as a simple linear combination of specular and diffuse reflection~\cite{xu_hyperthermal_2025, murray_experiment_2017, minton, poovathingal_experiments_2016}.
Consequently, a variety of models have been developed, such as the \gls{cl} model~\cite{cercignani_kinetic_1971} and the Washboard model~\cite{tully_washboard_1990}.
More recent work has introduced several extensions of the Washboard model.
This includes an extension of the hybrid Washboard-\gls{cl} formulation that incorporates a velocity dependent corrugation factor~\cite{park_corrugated_2025}, as well as generalized roughness models that account for anisotropic surfaces~\cite{jorge_generalized_roughness_2025}.
Another approach applies electromagnetic wave theory to represent roughness‑induced macroscopic effects in the scattering distribution~\cite{anton_wave_2025}.
All of these models use predefined functional forms, limiting the shape of the resulting scattering distribution.
Furthermore, they depend on accommodation coefficients or other parameters, which are generally treated as constant.
These parameters are difficult to determine and depend on a variety of factors, such as particle velocity, angle of incidence, surface temperature, and material composition~\cite{liang_parameter-free_2021, mehta_comparing_2014}.

An alternative to classical scattering models is to employ machine learning techniques that learn the scattering distribution directly from \gls{md} simulations.
\Gls{md} resolves every individual impact with atomistic precision, including full momentum and energy transfer at the collision interface.
As a result, machine learning models can extract the underlying scattering behavior without relying on predefined functional forms or accommodation coefficients.
This enables them to capture velocity-dependent effects and multi-modal distributions that simple parametric kernels cannot represent.
Such capabilities were demonstrated in recent work by the authors in~\cite{Schuette2025} for \gls{vleo} conditions.
However, this model was trained exclusively on hypersonic impacts ($6\,000\,$-$\,10\,000\,\text{m/s}$) and does not generalize to lower velocities.
This limitation becomes critical in multi-collision scenarios.
Particles entering \gls{abep} intakes at orbital speeds undergo successive wall impacts, decelerating through $4\,000\,\text{m/s}$, $2\,000\,\text{m/s}$, down to thermal velocities.
By the second, third, or fourth collisions, the incident velocities are entirely outside the training regime.
Similarly, the surface roughness resulting from manufacturing or erosion induces multiple bounces with progressive energy loss.
Consequently, for \gls{vleo} applications, both hypersonic non-equilibrium scattering and thermal (near-)equilibrium scattering are critically important.
Capturing these regimes accurately requires scattering kernels that remain valid across the entire thermal-to-hypersonic velocity spectrum.
Additionally, the original model did not enforce physical constraints like detailed balance, which can lead to thermodynamic inconsistencies in long-duration simulations.

This work addresses both limitations through a physics-informed conditional normalizing flow architecture trained on expanded \gls{md} data that cover the full velocity range from thermal ($\approx$$1\,000\,\text{m/s}$) to orbital speeds ($7\,800\,\text{m/s}$).
We incorporate a detailed balance loss term to ensure thermodynamic consistency while maintaining high fidelity to the underlying \gls{md} simulations.
The resulting model is compared to the previously used \gls{cvae} approach~\cite{Schuette2025}.
This framework enables accurate \gls{dsmc} predictions for complex \gls{vleo} applications, including \gls{abep} intake optimization and realistic surface roughness modeling.
However, the model is not limited to \gls{vleo} applications, but can be applied to any rarefied gas dynamics problem ranging from thermal to hypersonic velocities.
In addition to the application to \gls{dsmc}, the models are also applicable to \gls{tpmc}.
Furthermore, the aerodynamic coefficients obtained from this model can be integrated into panel methods such as ADBSat~\cite{sinpetru_adbsat_2022}, allowing accurate aerodynamic predictions for satellite design and mission planning in \gls{vleo}.

% ----------------------------------------------------------------------------------------------- %
% Methods
% ----------------------------------------------------------------------------------------------- %
\section{Methods} \label{sec:methods}

% Methods - Computational Methods
\subsection{Computational Methods} \label{subsec:computational_methods}

% Methods - Computational Methods - DSMC
\subsubsection{Direct Simulation Monte Carlo} \label{subsubsec:dsmc}

In \gls{vleo}, the mean free path of atmospheric particles exceeds typical satellite dimensions, resulting in a high Knudsen number ($\text{Kn} > 10$) flow regime.
Consequently, the continuum assumption underlying traditional fluid dynamics breaks down, and the gas exhibits non-equilibrium behavior.
The \gls{dsmc} method~\cite{bird_molecular_1994} addresses this regime by representing gas as discrete simulation particles.
It models the gas flow by tracking the trajectories of these particles, which represent a weighted number of real gas molecules or atoms.
These simulation particles approximate the distribution function $f$ in \gls{dsmc}.
The Boltzmann equation specifies how the particle distribution function $f(\bm{x}, \bm{v}, t)$ evolves in phase space.
Its general form is
\begin{equation} \label{eq:boltzmann}
    \frac{\partial f}{\partial t} + \bm{v} \cdot \frac{\partial f}{\partial \bm{x}} + \frac{\bm{F}}{m} \cdot \frac{\partial f}{\partial \bm{v}} = \left( \frac{\delta f}{\delta t} \right)_{\text{coll}}
\end{equation}
where $\bm{x}$ and $\bm{v}$ are the position and velocity vectors, $t$ is time, $\bm{F}$ is the external force acting on the particles, and $m$ is the particle mass.
The three terms on the left side represent the temporal variation, the advection in physical space due to particle motion, and changes in velocity induced by external forces $\bm{F}$.
The right-hand side accounts for momentum and energy-exchange during collisions between particles and governs the relaxation of the system toward its equilibrium distribution.
Therefore, the \gls{dsmc} method simulates the movement of particles through the computational domain and their interactions with each other and with boundaries, to approximate the Boltzmann equation.
For the interaction processes, probabilistic models are used.
Electromagnetic forces are not included in standard \gls{dsmc} formulations~\cite{bird_molecular_1994, nejad_scattering_2023}.

% Gas-surface interactions in DSMC
In \gls{vleo} applications, the \glspl{gsi} are of particular importance, as collisions between particles hardly occur in the free molecular flow regime.
As a result, the aerodynamic forces on the satellite are primarily determined by the interactions of impinging particles with the surface of the satellite.
The scattering of particles on the surface is modeled in \gls{dsmc} using scattering kernels.
It describes how an incident velocity distribution $f(\vecvi)$ is mapped into the reflected velocity distribution $f(\vecvr)$.
This relationship is expressed in equation~\eqref{eq:scattering_kernel}~\cite{nejad_scattering_2023, cercignani_kinetic_1971}:
\begin{equation}
    v_{\text{n,r}} \, f\left( \bm{v}_{\text{r}} \right) = \int_{v_{\text{n,i}}<0} |v_{\text{n,i}}| \, \skernel{\vecvi}{\vecvr} \, f\left( \vecvi \right) \, \mathrm{d} \vecvi
    \label{eq:scattering_kernel}
\end{equation}
Here, $v_{\text{n,r}}$ and $v_{\text{n,i}}$ denote the normal components of the reflected and incident velocities.
The kernel $\skernel{\vecvi}{\vecvr}$ acts as a conditional probability density, specifying the likelihood that a particle approaching with velocity $\vecvi$ is scattered to a state with velocity $\vecvr$.

To be physically valid, such a kernel must satisfy three fundamental conditions~\cite{cercignani_kinetic_1971,cercignani_boltzmann_1988}:
\begin{enumerate}
    \item Non-negativity requirement
    \begin{equation} \label{eq:non_negative_req}
        \skernel{\vecvi}{\vecvr} \geq 0
    \end{equation}
    The non-negativity requirement states that the scattering kernel must assign non-negative probabilities to all possible reflected velocities for every incident velocity.
    This ensures that the kernel represents a valid probability distribution as the probabilities cannot be negative.

    \item Normalization requirement
    \begin{equation} \label{eq:normalization_req}
        \int_{\vecvr \cdot \bm{n} > 0} \skernel{\vecvi}{\vecvr} \, \text{d}\vecvr = 1 \quad\quad \vecvi \cdot \bm{n} < 0
    \end{equation}
    The normalization requirement states that for every incident velocity with $\vecvi \cdot \bm{n} <0$, the scattering kernel must define a proper probability distribution over all physically admissible reflected velocities.
    In other words, the kernel must assign all its probability masses to the half-space $\vecvr \cdot \bm{n} > 0$, and the total mass over that domain must be exactly one.

    \item \Gls{db} requirement
    \begin{equation} \label{eq:db_req}
        \begin{aligned}
            &\underbrace{\mathcal{M}\left(\vecvi\right) \left|\vecvi \cdot \bm{n}\right| \skernel{\vecvi}{\vecvr}}_{\text{\gls{db}}_{\text{fwd}}} \\
            = &\underbrace{\mathcal{M}\left(\vecvr\right) \left|\vecvr \cdot \bm{n}\right| \skernel{-\vecvr}{-\vecvi}}_{\text{\gls{db}}_{\text{bwd}}}
        \end{aligned}
    \end{equation}
    Here, $\mathcal{M}$ denotes the Maxwell distribution of the gas in equilibrium with the wall.
    The \gls{db} relation ensures consistency with equilibrium thermodynamics.
    It requires that the \gls{fwd} probability flux from $\vecvi$ to $\vecvr$ is equal to the \gls{bwd} flux from $-\vecvr$ to $-\vecvi$ when the gas is in equilibrium with the wall.
    This can be physically interpreted as following~\cite{cercignani_boltzmann_1988}.
    If a gas is in equilibrium with the wall temperature $T_{\text{w}}$, and therefore has a Maxwell distribution $\mathcal{M}$ at that temperature, then the number of particles scattering from a velocity range $(\vecvi, \vecvi +\text{d}\vecvi)$ into a range $(\vecvr, \vecvr + \text{d}\vecvr)$ is exactly equal to the number of particles scattering from $(-\vecvr, -\vecvr + \text{d}\vecvr)$ into $(-\vecvi, -\vecvi + \text{d}\vecvi)$.
    One consequence of the \gls{db} relation is that if the incident velocity distribution is a Maxwellian flux $f_{\text{i}} \sim |v_{\text{i,n}}| \mathcal{M}(\vecvi) $ at the wall temperature, then the reflected velocity distribution will also be a Maxwellian flux  at the same temperature~\cite{cercignani_boltzmann_1988}.
    % Violation of this condition would imply artificial creation or destruction of momentum or energy at the wall, leading to unphysical equilibrium behavior.
\end{enumerate}

% Scattering kernels
Building on these fundamental requirements, one of the most widely used scattering models in \gls{dsmc} is the classical Maxwell kernel~\cite{Maxwell_stresses_1879}.
It assumes that an incident particle undergoes specular or diffuse reflection, represented by
\begin{equation} \label{eq:Maxwell_kernel}
    \begin{aligned}
        \skernel{\vecvi}{\vecvr} &= \underbrace{(1 - \sigma) \delta(\vecvr - \vecvi)}_{\text{specular}} \\
        &+ \underbrace{\sigma \mathcal{M}(\vecvr, T_{\text{w}}) |v_{\text{n,r}}|}_{\text{diffuse}}.
    \end{aligned}
\end{equation}
The specular term represents mirror-like reflection, and the diffuse term describes the re-emission of particles following a Maxwell flux distribution $\MaxwellFluxRe$ at the wall temperature $T_{\text{w}}$.
$\delta$ is the Dirac delta function and $\sigma$ the accommodation coefficient that quantifies the relative contributions of specular and diffuse reflections: $\sigma = 1$ implies purely diffuse behavior, while $\sigma = 0$ indicates purely specular reflection~\cite{Maxwell_stresses_1879, livadiotti_review_2020}.
Although it is often treated as a constant, $\sigma$ depends on multiple factors, including particle energy, angle of incidence, surface temperature, and material composition.
These dependencies are not well characterized and the accommodation coefficient is typically not known~\cite{liang_parameter-free_2021, mehta_comparing_2014}.
Moreover, molecular-beam experiments have demonstrated that real \glspl{gsi} cannot be represented as a simple linear combination of specular and diffuse components~\cite{xu_hyperthermal_2025, murray_experiment_2017, minton, poovathingal_experiments_2016}.
This has motivated the development of more sophisticated scattering models, such as the \gls{cl} kernel~\cite{cercignani_kinetic_1971} and the Washboard model~\cite{tully_washboard_1990}.
However, these approaches still rely on uncertain accommodation parameters and predefined functional forms, limiting their ability to capture complex scattering behavior.
Recently, a data-driven approach was proposed that combines \gls{md} simulations with a \gls{cvae} to learn the scattering kernel directly from microscopic simulation data without relying on predefined functional forms and unknown accommodation coefficients~\cite{Schuette2025}.
This method showed promising results in capturing the scattering behavior at hypersonic velocities, but is currently limited to single impacts.

% Methods - Computational Methods - MD
\subsubsection{Molecular Dynamics} \label{subsubsec:md}

With \gls{md} simulations, detailed information can be obtained on the physical processes governing \glspl{gsi}.
\Gls{md} describes the trajectories of gas particles and surface atoms by solving Newton's second law of motion,
\begin{equation}
    \label{eq:newton}
    m_{j} \frac{\text{d}^2 \bm{x}_j}{\text{d}t^2} = \bm{F}_{j},
\end{equation}
where $m_{j}$ is the mass of particle $j$, $\bm{x}_j$ is its position, and $\bm{F}_{j}$ is the force acting on it.
The interatomic forces acting on each particle are derived from a potential energy function $U$, by following relationship
\begin{equation}
    \bm{F}_{j} = - \nabla_{j} U(\bm{x}_1, \bm{x}_2, \dots, \bm{x}_N),
\end{equation}
where $\bm{x}_1, \bm{x}_2, \dots, \bm{x}_N$ denote the positions of all $N$ particles in the system.
This potential captures forces such as attraction, repulsion, and bonding between atoms, allowing for the prediction of system behavior at the atomic level.
Several strategies exist to calculate the potential energy function $U$.
Ab initio methods, such as those based on \gls{dft}, provide highly accurate descriptions by solving the Schrödinger equation.
They are computationally demanding, but remain the method of choice whenever high accuracy is required.
In contrast, parameterized empirical potentials offer a far more efficient alternative.
Their parameters are calibrated against experimental or ab initio data, allowing the resulting interaction models to support simulations at comparatively low computational expense\cite{allen_md_2017,chen_molecular_2023,nejad_scattering_2023}.

Although \gls{md} simulations offer atomistic detail on gas-surface interactions, their high computational cost restricts them to small systems and short timescales.
Nevertheless, the microscopic information obtained from \gls{md} simulations is crucial to build a data-driven scattering kernel for \gls{dsmc} simulations.
Therefore, specific sets of incident particles $\bm{v}_{\text{i}}$ are simulated using \gls{md} simulations to derive the corresponding reflected velocity distributions $P(\vecvr | \vecvi)$.
These conditional distributions can then be used to train a generative machine learning model, such as a \gls{cvae}, to predict the reflected velocity distribution for any given incident velocity $\vecvi$.
The resulting machine learning model can then be used as a scattering kernel in \gls{dsmc} simulations.

% Methods - Data-Driven Model
\subsection{Data-Driven Scattering Model} \label{subsec:scattering_model}

To construct a data-driven scattering model, we use the approach proposed by~\citet{Schuette2025}, which combines \gls{md} simulations with a generative machine learning model to learn the scattering kernel directly from microscopic simulation data without relying on predefined functional forms and unknown accommodation coefficients.

\begin{figure}[tb!]
    \centering
    \includegraphics{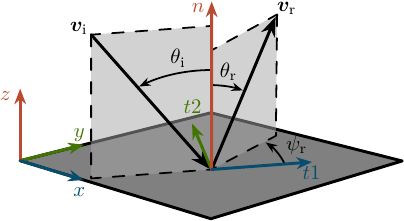}
    %\inputtikz{figures}{cos_wall_scattering}
    \caption{Global frame of reference ($x$, $y$, $z$) and local scattering frame of reference ($t1$, $t2$, $n$)~\cite{Schuette2025}.}
    \label{fig:cos}
\end{figure}

The first step is to obtain the required scattering datasets from \gls{md} simulations.
To keep the number of incident velocity vectors $\vecvi$ that must be simulated at a manageable level, the scattering process is assumed to be independent of the incident azimuthal angle $\psi_{\text{i}}$.
This implies that the surface interaction does not vary with rotations around the surface normal.
Such an approximation is well justified for isotropic surfaces, which is typical for amorphous surface structures, where scattering exhibits uniform properties in all directions.
In contrast, anisotropic surfaces, including crystalline substrates or engineered textures with grooves or patterned coatings, may violate this assumption because directional features can alter the scattering response.
Under this assumption, the scattering is performed in the local scattering frame of reference ($t1$, $t2$, $n$), as shown in Figure~\ref{fig:cos}.
In this reference frame, the global coordinate system ($x$, $y$, $z$) is rotated around the surface normal $n$ such that the incident velocity component in the $t2$-direction is zero.
Consequently, only different incident velocity magnitudes $|v_{\text{i}}|$ and polar angles $\theta_{\text{i}}$ should be selected for data generation, significantly reducing the computational cost of \gls{md} simulations~\cite{Schuette2025}.
Each selected incident velocity vector $\vecvi$ is impacted onto the surface several times to obtain the corresponding reflected velocity distribution $p(\vecvr | \vecvi)$.

The resulting datasets are then used to train and validate generative machine learning models that learn to approximate the reflected velocity distribution $p(\vecvr | \vecvi)$ for any given incident velocity $\vecvi$.
Two architectures are examined in this study: the already mentioned \gls{cvae} and a conditional normalizing flow model, both are introduced in the subsequent Sections.

% Methods - Data-Based Model - cVAE
\subsubsection{Conditional Variational Autoencoder} \label{subsubsec:cvae}

\begin{figure}
    \centering
    \includegraphics{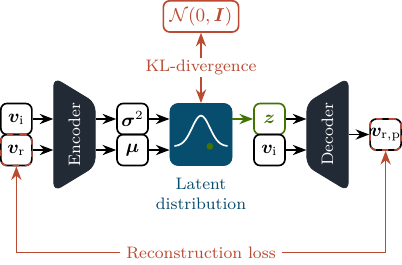}
    %\inputtikz{figures}{cvae_model}
    \caption{Schematic of the \gls{cvae} model architecture~\cite{Schuette2025}.}
    \label{fig:cvae_model}
\end{figure}

A \gls{cvae} can learn to generate new data points that are similar to a given dataset, while also allowing for conditioning on additional information.
In~\citet{Schuette2025} a detailed description of the \gls{cvae} architecture and training procedure is given, so here only a brief overview is provided.
The \gls{cvae} consists of an encoder and a decoder network with a latent distribution in between (see Figure~\ref{fig:cvae_model}).
The encoder takes the reflected velocity $\vecvr$ and the incident velocity $\vecvi$ as input and parametrizes the latent distribution, which is typically a diagonal multivariate normal distribution $\mathcal{N}(\bm{\mu}, \bm{\sigma}^2)$.
From this distribution a latent vector $\bm{z}$ is sampled and passed to the decoder, which takes $\bm{z}$ and $\vecvi$ as input and outputs a reconstructed reflected velocity $\vecvrp$.
The model is trained to minimize the reconstruction error between $\vecvr$ and $\vecvrp $ and to regularize the latent space to follow a standard normal distribution $\mathcal{N}(0, \bm{I})$ by adding a \gls{kl} divergence term to the loss function.
After training, the decoder can be used as a generative model for reflected velocities conditioned on incident velocity $\vecvi$.
However, the \gls{cvae} does not allow direct evaluation of the probability density value of the scattering kernel $\skernel{\vecvi}{\vecvr}$.
Instead, it implicitly defines the kernel through a conditional generative process.
A latent variable $z \sim \mathcal{N}(0, \bm{I})$ is deterministically assigned to a reflected velocity $\vecvr$ by passing it through the decoder network together with the incident velocity $\vecvi$.

For the predicted reflected velocity $\vecvr$ the normal velocity component $v_{\text{n}}$ must be positive, because the particles are reflected from the surface and cannot penetrate it.
To enforce this constraint, the Softplus activation function~\cite{dugas_softplus_2000}
\begin{equation} \label{eq:softplus}
    \text{Softplus}(x; \beta) = \frac{1}{\beta} \log\left(1 + \exp(\beta x)\right)
\end{equation}
is applied to the normal velocity component of the decoder's output.

% Non-negativity requirement
The \gls{cvae} is used purely as a conditional generative model and does not output the scattering kernel $\skernel{\vecvi}{\vecvr}$ directly.
Instead, the decoder defines the scattering kernel implicitly as the pushforward of the latent normal distribution through a deterministic neural mapping.
Push-forwards of probability measures are themselves valid probability measures and, therefore, non-negative everywhere on their support.
The non-negativity requirement~\eqref{eq:non_negative_req} is thus fulfilled.

% Normalization requirement
Due to the fact that the \gls{cvae} decoder is a deterministic mapping of a latent normal distribution, it induces a normalized conditional distribution $p(\vecvrp| \vecvi)$, since any distribution obtained as the push-forward of a normalized latent distribution is itself normalized.
The Softplus activation applied to the normal component of the decoder's output ensures $\vecvr \cdot \bm{n}>0$ for all generated reflected velocities.
Thus, the decoder maps all latent samples into the physically admissible half-space.
Since the entire probability mass of the induced distribution lies in $\vecvr \cdot \bm{n}>0$ and the latent distribution is normalized, the resulting conditional distribution is automatically normalized over the correct physical domain.
Therefore, the normalization requirement~\eqref{eq:normalization_req} is satisfied.

% Detailed balance requirement
As already mentioned, the conditional probability density $p(\vecvr | \vecvi)$ associated with the \gls{cvae} decoder is not available in closed form.
The decoder provides only a sampling mechanism, i.e. a push-forward of the latent prior through a deterministic neural mapping, rather than an explicit expression for the corresponding probability density function.
As a consequence, neither side of the detailed-balance relation~\eqref{eq:db_req} can be evaluated directly.
Without an explicit density, it is impossible to verify whether the learned kernel satisfies detailed balance.
Moreover, since the architecture does not impose any symmetry constraints relating forward and backward scattering, it is unlikely that the \gls{cvae} satisfies the detailed-balance requirement except by coincidence or approximation inherited from the training data.

% Methods - Data-Based Model - cNormalizing Flow
\subsubsection{Conditional Normalizing Flow} \label{subsubsec:crealnvp}

Normalizing flows learn to transform a simple base distribution, typically a standard normal distribution $\mathcal{N}(0, \bm{I})$, into a complex target distribution by applying a sequence of $K$ invertible, differentiable mappings $f_1, \dots, f_K$.
The complete transformation $f = f_K \circ \cdots \circ f_1$ is itself invertible, since the composition of invertible functions is also invertible.
For a composition of invertible functions, the inverse can be computed by applying the inverses of the individual functions in reverse order, i.e.~\cite{dinh2015nice, rezende2016variational}
\begin{equation}
    f^{-1} = (f_K \circ \cdots \circ f_1)^{-1} = f_1^{-1} \circ \cdots \circ f_K^{-1}.
\end{equation}
This allows both the forward pass (training) and the inverse pass (sampling) to be computed exactly and efficiently, without any approximation.

One way of constructing a normalizing flow is to use affine coupling layers, as done in \gls{realnvp} model~\cite{dinh2017realnvp}.
This architecture can be extended to a conditional version, the \gls{crealnvp}~\cite{winkler2023conditionalnormalizingflows}, by conditioning the transformations on additional input data, in our case the incident velocity $\vecvi$.
\begin{figure*}[htb]
    \centering
    \includegraphics{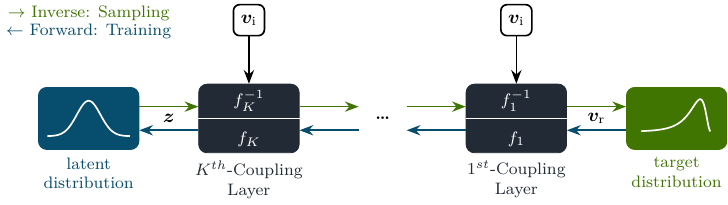}
    %\inputtikz{figures}{normalizing_flow}
    \caption{Schematic of the  \gls{crealnvp} architecture.}
    \label{fig: crealnvp_model}
\end{figure*}
During training, the forward pass $f: \vecvr \mapsto \bm{z}$ maps the data (reflected velocity $\vecvr$) to the latent space to evaluate the likelihood.
\begin{equation}
    \bm{z} = f_K \circ \cdots \circ f_1(\vecvr, \vecvi)
\end{equation}
During sampling, the inverse pass $f^{-1}: \bm{z} \mapsto \vecvr$ transforms a point $\bm{z}$ sampled from the base distribution into a new data point (a generated reflected velocity $\vecvr$).
\begin{equation}
    \vecvr = f_1^{-1} \circ \cdots \circ f_K^{-1}(\bm{z}, \vecvi)
\end{equation}
The schematic of a normalizing flow architecture is shown in Figure~\ref{fig: crealnvp_model} with the forward and inverse pass.

\begin{figure}[tb!]
    \begin{subfigure}{\columnwidth}
        \centering
        \includegraphics{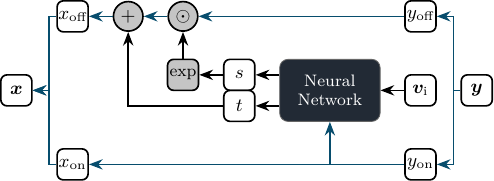}
        %\inputtikz{figures}{coupling_layer_forward}
        \caption{Forward pass $f_k(\bm{y}; \vecvi)$}
        \label{fig:coupling_layer_forward}
    \end{subfigure}
    \begin{subfigure}{\columnwidth}
        \centering
        \includegraphics{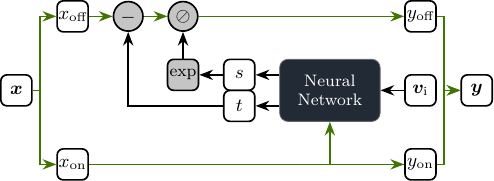}
        %\inputtikz{figures}{coupling_layer_inverse}
        \caption{Inverse pass $f^{-1}_k(\bm{x}; \vecvi)$}
        \label{fig:coupling_layer_inverse}
    \end{subfigure}
    \caption{Schematic of the forward and inverse pass through coupling layer $k$ in \gls{crealnvp}.}
\end{figure}

A \gls{crealnvp} model consists of a sequence of $K$ conditional affine coupling layers $f_k$.
Each layer splits the input into two parts.
The ``on'' part $\bm{y}_{\text{on}}$ remains unchanged, while the ``off'' part $\bm{y}_{\text{off}}$ is transformed by an affine transformation.
The on and off parts can be any combination of the input dimensions, for example, the first two velocity components can be on while the last component is off, or vice versa.
Active dimensions in a given layer are controlled by a binary mask that can change from layer to layer to ensure that all dimensions interact across the layers.
In the forward pass, the $k$-th coupling layer $f_k$ transforms the input $\bm{y}$ to $\bm{x}$ as follows
\begin{equation}
    \begin{aligned}
        \bm{x}_{\text{off}} &= \bm{y}_{\text{off}} \odot \exp\bigl(s(\bm{y}_{\text{on}}, \vecvi)\bigr) + t(\bm{y}_{\text{on}}, \vecvi), \\
        \bm{x}_{\text{on}}  &= \bm{y}_{\text{on}}
    \end{aligned}
\end{equation}
where $\odot$ denotes element-wise multiplication.
The scaling $s$ and translating $t$ parameters of the affine transformation are predicted by a neural network that takes the preserved part and the conditioning variable $\vecvi$ as input.
Inverse mapping $f_k^{-1}$ (used during sampling) is obtained analytically by inverting the affine transform:
\begin{equation}
    \begin{aligned}
        \bm{y}_{\text{off}} &= \bigl(\bm{x}_{\text{off}} - t(\bm{x}_{\text{on}}, \vecvi)\bigr) \oslash \exp\bigl(s(\bm{x}_{\text{on}}, \vecvi)\bigr), \\
        \bm{y}_{\text{on}}  &= \bm{x}_{\text{on}}
    \end{aligned}
\end{equation}
Importantly, the same neural network (i.e., the same weights) is used to compute $s$ and $t$ for both directions.
At the end of a coupling layer, the on and off parts ($\bm{x}_{\text{off}}$ and $\bm{x}_{\text{on}}$) are concatenated to get the output $\bm{x}$ of the layer.
In Figure~\ref{fig:coupling_layer_forward} the forward pass through a coupling layer is shown, where the input $\bm{y}$ is transformed into $\bm{x}$.
The inverse pass through the same coupling layer is shown in Figure~\ref{fig:coupling_layer_inverse}, where the input $\bm{x}$ is transformed back to $\bm{y}$.
Every coupling layer is analytically invertible, allowing both the forward pass (data to latent space) and the inverse pass (latent space to data) to be exact and efficient.

The complete forward transformation $f: \vecvr \mapsto \bm{z}$ maps the reflected velocity $\vecvr$ to a latent variable $\bm{z}$, which follows the base distribution $p_Z(\bm{z})$.
Because the transformation of each coupling layer is invertible, the exact conditional log-probability follows from the change-of-variables formula:
\begin{equation} \label{eq:log_prob}
    \begin{aligned}
        \log p(\vecvr | \vecvi) &= \log p_{Z}(f(\vecvr , \vecvi)) \\
        &+ \sum_{k=1}^{K} \log \left| \det \left( \frac{\partial f_k}{\partial \bm{x}^{(k)}} \right) \right|
    \end{aligned}
\end{equation}
The first term evaluates how likely the mapped latent variable $\bm{z} = f(\vecvr, \vecvi)$ is under the base distribution.
The second term, the logarithmic absolute determinant of the Jacobian, accounts for the change in volume during the transformation.
Since the Jacobian of each affine coupling layer is triangular, the determinant can be efficiently computed as the product of the diagonal elements.
These diagonal elements correspond to the scale outputs $s_l^{(k)}$ of the off dimensions $l \in \text{off}^{(k)}$ at each layer $k$, rendering the log-probability exact and efficient to evaluate:
\begin{equation}
    \begin{aligned}
        \log p(\vecvr | \vecvi) &= \log p_{Z}(f(\vecvr , \vecvi)) \\
        &+ \sum_{k=1}^{K} \sum_{l\in\text{off}^{(k)}} s_l^{(k)}(\bm{x}_{\text{on}}^{(k)}, \vecvi)
    \end{aligned}
\end{equation}

During training, the exact conditional log-likelihood over the dataset $\left\{(\bm{v}_{\text{r},j}, \bm{v}_{\text{i},j}) \right\}$ is maximized, which is equivalent to minimizing the \gls{nll} loss:
\begin{equation} \label{eq:nll_loss}
    \mathcal{L}_\text{\gls{nll}} = -\frac{1}{N} \sum_{j=1}^{N} \log p(\bm{v}_{\text{r},j} | \bm{v}_{\text{i},j})
\end{equation}

The \gls{crealnvp} natively operates on the full three-dimensional velocity space $\bm{v} \in \mathbb{R}^3$ without coordinate restrictions.
The affine coupling layers transform all three velocity components jointly, across all the stacked layers.
However, in case of surface scattering, the normal velocity component $v_{\text{n}}$ is physically constrained to be strictly positive.
Since the \gls{crealnvp} is defined on $\mathbb{R}^3$, it can produce samples with negative normal velocity components, which are unphysical.
A clean solution is to apply a preprocessing transformation before passing data to the \gls{crealnvp}.
Such a transformation must have a few key properties.
It must be differentiable, so the Jacobian can be computed, and it must be invertible (bijective) to ensure there is a unique mapping between the model's output space and the physical velocity space.
The Softplus function~\eqref{eq:softplus} fulfills these requirements.
With the inverse Softplus, the normal velocity component $v_{\text{n}}$ can be transformed into an unconstrained variable $\tilde{v}_{\text{n}}$ as follows:
\begin{equation}
    \tilde{v}_{\text{n}} = \text{Softplus}^{-1}(v_{\text{n}}; \beta) = \frac{1}{\beta} \log\left(\exp(\beta v_{\text{n}}) - 1\right)
\end{equation}
This maps $v_{\text{n}} \in (0, \infty)$ bijectively to $\tilde{v}_{\text{n}} \in \mathbb{R}$, on which the \gls{crealnvp} can operate without constraints.
The inverse transformation,
\begin{equation}
    v_{\text{n}} = \text{Softplus}(\tilde{v}_{\text{n}}; \beta) = \frac{1}{\beta} \log\left(1 + \exp(\beta \tilde{v}_{\text{n}})\right)
\end{equation}
maps back to the strictly positive normal velocity space during sampling.
The Softplus preprocessing transformation does not have to be added to the \gls{nll} loss, because the transformation occurs before the data are passed to the \gls{crealnvp} and from the model's perspective the physical normal velocity constraint does not exist.
The model learns the unrestricted distribution over $\mathbb{R}^3$ of the tangential velocities $v_{\text{t1}}$, $v_{\text{t2}}$ and the transformed normal velocity $\tilde{v}_{\text{n}}$.
However, if one wants to evaluate the physical probability density of the reflected velocity samples, the Softplus transformation has to be considered in the change of variables formula, which adds an additional term to equation~\eqref{eq:log_prob}:
\begin{equation}
    \log \left| \frac{\text{d} \tilde{v}_{\text{n}} }{\text{d} v_{\text{n}} } \right| = - \log \left( \frac{\exp(\beta\tilde{v}_{\text{n}})}{1 + \exp(\beta\tilde{v}_{\text{n}})} \right)
\end{equation}

Another approach, to force positive normal velocity, would be to add the Softplus transformation to the model architecture itself.
The model takes raw, untransformed data as input, where the normal velocity component is strictly positive, and applies the Softplus transformation as the first layer of the model.
In the following, the subsequent coupling layers operate on transformed normal velocity $\tilde{v}_{\text{n}}$ and tangential velocities $v_{\text{t1}}$, $v_{\text{t2}}$, which now lie entirely in $\mathbb{R}^3$.
The crucial difference is that the Jacobian of this initial transformation layer must be included in the \gls{nll} loss during training.

% Non-negativity requirement
As a first step in verifying that the \gls{crealnvp}-based scattering kernel meets the kernel requirements, we examine non-negativity.
The \gls{crealnvp} based scattering kernel is defined through an explicit conditional probability density~\eqref{eq:log_prob}.
This density $p(\vecvr|\vecvi)$ is defined by the non-negative base distribution and the absolute value of the Jacobian determinant of the inverse transformation, which is also non-negative.
Since both terms are non-negative, the resulting conditional probability density is non-negative everywhere on its support, thus satisfying the non-negativity requirement~\eqref{eq:non_negative_req}.

% Normalization requirement
Normalization is preserved by construction in the \gls{crealnvp} architecture.
The induced conditional density integrates to one over its entire support, because the transformation between latent space and velocity space is bijective and differentiable.
Therefore, a normalizing flow preserves the probability mass by construction.
When the normal component of the reflected velocity is constrained to satisfy $ \vecvr \cdot \bm{n} $, the mapping sends the entire latent distribution into the physically admissible half-space $ \vecvr \cdot \bm{n} $.
As a result, the conditional distribution is automatically normalized over the correct physical domain and the normalization requirement is fulfilled.

% Detailed balance requirement
Although the \gls{crealnvp} architecture provides an explicit and tractable expression for the conditional density, it does not impose any symmetry constraints that would enforce the \gls{db} relation.
The model may approximately capture it if the training data itself satisfies detailed balance, but there is no architectural constraint that enforces it.
Compared to the \gls{cvae}, the \gls{crealnvp} has the advantage of directly evaluating the detailed balance relation, since it gives the exact conditional probability density function value for any reflected velocity sample.
This also allows to add a detailed balance loss term to the training loss, which can help to approximate the detailed balance requirement.
\begin{equation} \label{eq:loss_db}
    \begin{aligned}
        \mathcal{L}_{\text{\gls{db}}} &= \frac{1}{N} \sum_{j=1}^{N} \left( \mathcal{M}\left(\vecvi\right) \left|\vecvi \cdot \bm{n}\right| \skernel{\vecvi}{\vecvr} \right. \\
        & \left. - \mathcal{M}\left(\vecvr\right) \left|\vecvr \cdot \bm{n}\right| \skernel{-\vecvr}{-\vecvi} \right)^2
    \end{aligned}
\end{equation}
The detailed balance loss has to be evaluated in the physical velocity space.
Therefore, if the Softplus transformation is used as a preprocessing step, the log-probability of the model has to be evaluated in the transformed space, and then the additional term from the Softplus transformation has to be added to get the correct log-probability in the physical space.
However, a loss is just a soft constraint and it does not force the model to satisfy detailed balance exactly.

% ----------------------------------------------------------------------------------------------- %
% Training and Validation Data
% ----------------------------------------------------------------------------------------------- %
\section{Training and Validation Data} \label{sec:training_data}
For both models the same training and validation data is used, which are described in more detail in the following Sections.
The dataset comprises two groups: equilibrium data and non-equilibrium data.
The equilibrium data ensure that both models are informed about the correct equilibrium behavior, a fundamental requirement for any scattering kernel.
The non-equilibrium data train the models to reproduce the scattering behavior across a wide range of incident velocities, from slow near-thermal conditions to hypersonic impacts.
This should allow the models to be applicable across a broad spectrum of applications from near-equilibrium conditions to highly non-equilibrium scenarios, such as those encountered in \gls{vleo}.

\subsection{Equilibrium Data} \label{subsubsec:equilibrium_data}
As mentioned previously, in equilibrium an incident Maxwell flux distribution of particles with wall temperature $\Tw$ is reflected as a Maxwell flux distribution with the same temperature.
Therefore, the equilibrium data consists of incident and reflected velocity pairs $\left\{(\vecvr, \vecvi)\right\}$, where both $\vecvr$ and $\vecvi$ are sampled from a Maxwell flux distribution $\MaxwellFlux$ at a temperature of $300\,\text{K}$.
For each validation and training set, $5\,000$ velocity pairs are sampled, resulting in a total of $10\,000$ velocity pairs for the equilibrium data.

\subsection{Non-Equilibrium Data} \label{subsubsec:non_equilibrium_data}

Details about the data generation of the non-equilibrium data are described in~\citet{Schuette2025}, and only a brief overview is summarized here.

The previously developed free stream \gls{cvae} model~\cite{Schuette2025} used five free stream velocity magnitudes between $6\,000$ and $10\,000\,\text{m/s}$.
These velocity magnitudes cover different atmosphere temperatures and orbit velocities in \gls{vleo}.
Nine incident angles between $0\,^{\circ}$ and $80\,^{\circ}$ were simulated for each magnitude, resulting in 45 different incident velocity vectors $\vecvi$.
Each incident condition was impacted $5\,000$ times on the surface and the corresponding reflected velocities $\vecvr$ were recorded.
This procedure produces 45 conditional reflected velocity distributions $p(\vecvr | \vecvi)$, one for each incident velocity vector.

In order to extend the applicability of such a data-driven scattering model, additional non-equilibrium data is generated for slower incident velocity magnitudes.
To cover the full velocity spectrum from thermal to hypersonic conditions, incident magnitudes of $2\,000$, $3\,000$ and $5\,000\,\text{m/s}$ are simulated as well.
Each of these three additional velocities is combined with the same nine incident angles, producing $27$ additional incident velocity vectors and their corresponding reflected velocity distributions.

As surface material, aluminum oxide $\text{Al}_2\text{O}_3$ is chosen.
This choice reflects the fact that aluminum is widely used in satellite structures and surfaces in \gls{vleo} are continuously exposed to atomic oxygen, which leads to the formation of a native oxide layer.
To represent this realistic surface condition, the $\text{Al}_2\text{O}_3$  structure used for the oxygen impingement simulations is modeled as an amorphous block.

% MD simulations
The \gls{md} simulations are carried out using the \gls{imd} package, which is developed at the institute for Functional Matter and Quantum Technologies at the University of Stuttgart~\cite{roth_imp_2019}.
\Gls{imd} is designed for large-scale \gls{md} studies in materials science and supports a broad spectrum of interaction models, including long-range interactions and polarizability, both essential for the present work.
Atomic interactions in the simulations are described by classical analytical potentials based on the Tangney-Scandolo model~\cite{tangney_potentials_2002}.
In this potential, atomic oxygen is polarizable, and their dipole moments are calculated self-consistently from the local electric field originating from surrounding charges and dipoles.
Short-range interactions are represented through a Morse-Stretch potential.

In order to satisfy the Nyquist criterion to ensure an accurate resolution of fast vibrational modes and numerical stability, a simulation time step of $0.1$ in the internal unit system is used.
The internal unit system is based on atomic-scale reference values for length, energy, and mass.
In physical units, this corresponds to a time step of approximately $1.018\,\text{fs}$.
Before impact simulations, the surface is equilibrated at a temperature of $300\,\text{K}$.
At the start of the simulation, the incident particle is placed at a distance of $>0.1\,\text{nm}$ from the surface to prevent premature interactions before the simulation begins.
In addition, the initial position of the incident particle is randomized to capture a diverse range of impact scenarios.

% ----------------------------------------------------------------------------------------------- %
% Model Construction and Training
% ----------------------------------------------------------------------------------------------- %
\section{Model Construction and Training} \label{sec:model_construction_and_training}
For constructing the \gls{cvae} and the \gls{crealnvp} models, the same three steps are followed: velocity scaling, data preprocessing, selection of hyperparameters like number of layers, neurons, activation functions, etc., and training of the model.
All these steps are performed separately for the \gls{cvae} and the \gls{crealnvp}, but the same training and validation data are used for both models.
For constructing the \gls{cvae} and the \gls{crealnvp}, we utilize the Keras library~\cite{chollet_keras_2015} with TensorFlow~\cite{tensorflow2015-whitepaper} as backend.
The training of the models, as well as the application of the models in Section~\ref{sec:application}, is performed with an AMD Ryzen 5 Pro 7530u CPU.
Therefore, the computational times mentioned in the following Sections are expected to be significantly reduced when using a GPU.
Additionally, with the use of a GPU a large hyperparameter search can be performed to further optimize the model architecture and training procedure, which is currently limited by the computational resources available.
Nevertheless, different architectures and training procedures were tested, and the configurations described in the following Sections exhibited good performance in terms of training and validation metrics.

\subsection{Velocity Scaling} \label{subsec:velocity_scaling}

\begin{figure*}[tb!]
    \centering
    \begin{minipage}{\textwidth}
        \centering
        \begin{minipage}[b]{0.32\textwidth}
            \centering
            \includegraphics{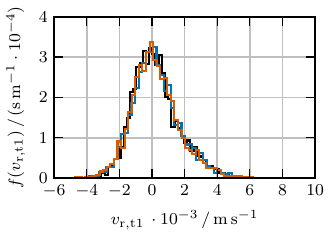}
            %\inputtikz{figures}{velocity_distribution_t1_different_Tw_7800_20}
        \end{minipage}\hfill%
        \begin{minipage}[b]{0.32\textwidth}
            \centering
            \includegraphics{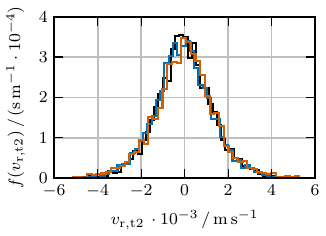}
            %\inputtikz{figures}{velocity_distribution_t2_different_Tw_7800_20}
        \end{minipage}\hfill%
        \begin{minipage}[b]{0.32\textwidth}
            \centering
            \includegraphics{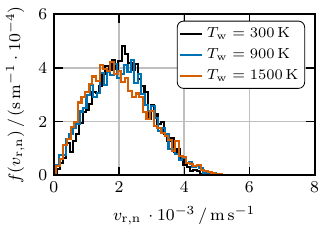}
            %\inputtikz{figures}{velocity_distribution_n_different_Tw_7800_20}
        \end{minipage}
        \vspace{-2mm}
        \subcaption{$|\bm{v}_{\text{i}}| = 7800.0 \, \text{m}/\text{s}$, $\theta_{\text{i}} = 20.0 \, ^\circ$}
    \end{minipage}
    \begin{minipage}{\textwidth}
        \centering
        \begin{minipage}[b]{0.32\textwidth}
            \centering
            \includegraphics{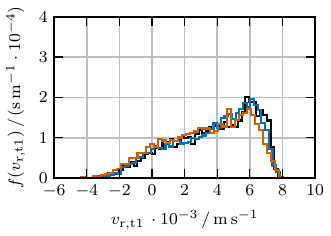}
            %\inputtikz{figures}{velocity_distribution_t1_different_Tw_7800_70}
        \end{minipage}\hfill%
        \begin{minipage}[b]{0.32\textwidth}
            \centering
            \includegraphics{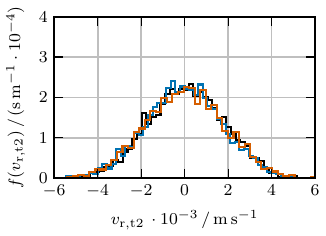}
            %\inputtikz{figures}{velocity_distribution_t2_different_Tw_7800_70}
        \end{minipage}\hfill%
        \begin{minipage}[b]{0.32\textwidth}
            \centering
            \includegraphics{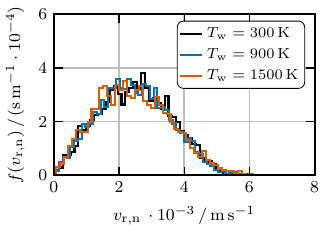}
            %\inputtikz{figures}{velocity_distribution_n_different_Tw_7800_70}
        \end{minipage}
        \vspace{-2mm}
        \subcaption{$|\bm{v}_{\text{i}}| = 7800.0 \, \text{m}/\text{s}$, $\theta_{\text{i}} = 70.0 \, ^\circ$}
    \end{minipage}

    \vspace{-2mm}
    \caption{Reflected velocity distributions of atomic oxygen of the \gls{md} data at different wall temperatures $\Tw$ for an incident velocity magnitude of $|\vecvi| = 7800.0\,\text{m/s}$ and varying polar angles $\theta_{\text{i}}$.}
    \label{fig:histograms_v_mag_7800.0_different_Tw_MD}
\end{figure*}

The reflected velocities in equilibrium and non-equilibrium are generated for a surface held at a wall temperature of $300\,\text{K}$.
To generalize across wall temperatures, a scaling procedure is applied to the velocities.
This procedure aims to eliminate the temperature dependence of the data.
For the equilibrium data, the temperature dependence is removed by normalizing velocities with the thermal velocity $v_{\text{th}}$
\begin{equation}
    v_{\text{th}} = \sqrt{\frac{2 k_B \Tw}{m}}.
\end{equation}
Here, $k_B$ denotes the Boltzmann constant, $\Tw$ the wall temperature and $m$ the particle mass.
This normalization follows from the equilibrium distributions of the tangential and normal velocity components.
These are described by a normal distribution $\mathcal{N}$ and Rayleigh distribution $\mathcal{R}$, respectively~\cite{garcia_maxwellian_2006}.
Both components can be expressed using $v_{\text{th}}$ as scale parameter:
\begin{align}
    v_{\mathcal{N}} &= \sqrt{\frac{1}{2}} v_{\text{th}} \mathfrak{R}_{\mathcal{N}} \quad \text{with } \mathfrak{R}_{\mathcal{N}} \sim \mathcal{N}(0, 1)\\
    v_{\mathcal{R}} &= \sqrt{\frac{1}{2}} v_{\text{th}} \mathfrak{R}_{\mathcal{R}} \quad \text{with } \mathfrak{R}_{\mathcal{R}} \sim \mathcal{R}(1)
\end{align}
Here, $\mathfrak{R}_{\mathcal{N}}$ and $\mathfrak{R}_{\mathcal{R}}$ denote standard normal $\mathcal{N}(0, 1)$ and unit Rayleigh $\mathcal{R}(1)$ random variables.
Normalization by $v_{\text{th}}$ renders the velocities non-dimensional.
This eliminates the dependence on wall temperature.
It also eliminates dependence on particle mass.
Consequently, the models only need to capture the standard normal distribution and the unit Rayleigh distribution.
Applying the inverse transformation to the model output reintroduces the wall-temperature and particle-mass dependence, recovering physical velocities.
The models can therefore be applied to arbitrary wall temperature $\Tw$ and particle mass $m$ without retraining in the equilibrium regime.

To assess how non-equilibrium scattering depends on wall temperature, reflected velocity distributions from \gls{md} simulations are shown for different wall temperatures $\Tw$ of $900\,\text{K}$ and $1500\,\text{K}$ in Figure~\ref{fig:histograms_v_mag_7800.0_different_Tw_MD}.
Relative to the $300\,\text{K}$ wall temperature, the reflected velocity distributions at higher wall temperatures are nearly identical.
Notably, $1500\,\text{K}$ is close to the melting temperature of aluminum oxide, which is $2326\,\text{K}$~\cite{crc_handbook_of_chemistry_and_physics}.
This suggests that the scattering behavior in this regime may be largely independent of $\Tw$ over a wide temperature range, up to near the material's melting point.
However, more data is needed to confirm this hypothesis.
For the purpose of this work the scattering behavior in the non-equilibrium regime is assumed to be independent of wall temperature.
A constant scaling factor is used based on the thermal velocity $v_{\text{th,300}}$ at $300\,\text{K}$.

\begin{figure*}[tb!]
    \centering
    \begin{subfigure}[t]{0.45\textwidth}
        \centering
        \includegraphics{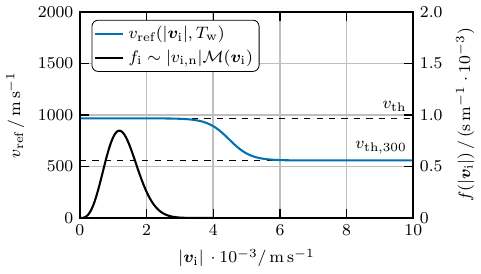}
        %\inputtikz{figures}{sigmoid_scaling_factor_900}
        \caption{$\Tw = 900\, \text{K}$}
        \label{fig:sigmoid_v_scaling_factor_900}
    \end{subfigure}
    \begin{subfigure}[t]{0.45\textwidth}
        \centering
        \includegraphics{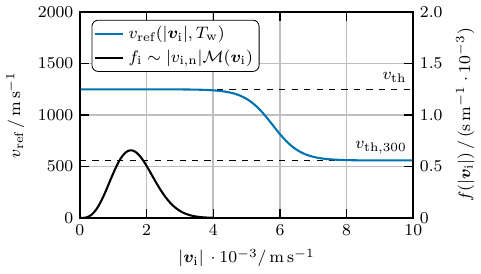}
        %\inputtikz{figures}{sigmoid_scaling_factor_1500}
        \caption{$\Tw = 1500\, \text{K}$}
        \label{fig:sigmoid_v_scaling_factor_1500}
    \end{subfigure}
    \caption{Velocity- and temperature-dependent reference velocity $v_{\text{ref}}$ for scaling, based on the sigmoid function $s$. The scaling converges to the thermal velocity $v_{\text{th}}$ in the equilibrium regime and to $v_{\text{th,300}}$ in the non-equilibrium regime (high incident velocities), as shown by comparison with the equilibrium distribution $f_{\text{i}}$ (Maxwell flux).}
    \label{fig:sigmoid_v_scaling_factor}
\end{figure*}

Between the constant plateaus of the equilibrium and non-equilibrium regimes lies a transition region.
Here, scattering behavior changes gradually from temperature-dependent to temperature-independent.
To describe the transition of the scaling factor between the two constant plateaus, a sigmoid function $\varsigma$ is used to smoothly interpolate between the two scaling regimes.
This results in the reference velocity $v_{\text{ref}}$ used for velocity scaling, defined as a function of the incident velocity magnitude $|\vecvi|$ and the wall temperature $\Tw$.
The scaling factor $v_{\text{ref}}$ is defined as:
\begin{align}
    v_{\text{ref}} &= \varsigma \, v_{\text{th,300}} + (1 - \varsigma) \, v_{\text{th}}\\
    \varsigma &= \frac{1}{1 + \exp(-a (b - 0.5))}\\
    \text{with} \quad b &= \frac{|\vecvi| - v_{\text{min}}}{v_{\text{max}} - v_{\text{min}}}
\end{align}

Here, $b$ denotes the normalized incident velocity magnitude.
It equals $0$ at the minimum incident velocity $v_{\text{min}}$ and $1$ at the maximum incident velocity $v_{\text{max}}$.
For $v_{\text{min}}$, the $97\,\%$ quantile of the incident Maxwell flux distribution at wall temperature $\Tw$ is used, which achieves an almost constant scaling factor equal to $v_{\text{th}}$ for incident velocities in the equilibrium regime.
From visual inspection of the incident velocity distributions and the reference velocity $v_{\text{ref}}$, $v_{\text{max}}$ is set to three times $v_{\text{min}}$, which ensures convergence to a constant value equal to $v_{\text{th,300}}$ for velocities in the non-equilibrium regime.
The steepness parameter $a$ is chosen such that $v_{\text{ref}}$ is $1\,\%$ below $v_{\text{th}}$ at $v_{\text{min}}$, and $1\,\%$ above $v_{\text{th,300}}$ at $v_{\text{max}}$, yielding a smooth transition in $v_{\text{ref}}$ while preserving the two plateaus corresponding to equilibrium and non-equilibrium behavior.
The resulting scaling factor $v_{\text{ref}}$ is shown in Figure~\ref{fig:sigmoid_v_scaling_factor_900} for a wall temperature of $\Tw = 900\,\text{K}$ and a wall temperature of $\Tw = 1500\,\text{K}$ in Figure~\ref{fig:sigmoid_v_scaling_factor_1500}.

While the scaling in equilibrium is exact, it should be emphasized that the proposed scaling is only an approximation outside thermal equilibrium.
However, the scaling with the temperature in non-equilibrium is reasonable to assume, since a higher $\Tw$ merely increases the kinetic energy of the wall atoms, while the interaction potential governing the \gls{md} simulations remains unchanged.
Generalization across species is more questionable in non-equilibrium, since the interaction potential in \gls{md} simulations differs for each species.
Consequently, this scaling approach cannot be expected to generalize across species outside thermal equilibrium.
For further validation, additional \gls{md} simulations at different wall temperatures are required to confirm the validity of this scaling approach.
The optimal approach for generalizing the model across wall temperatures would be training on a range of wall temperatures and using $\Tw$ as an additional model input.
Nevertheless, this would substantially increase the required training data and computational resources.
Therefore, the scaling technique presented here serves as a first step toward a generalized model for varying wall temperatures.

\subsection{\gls{cvae}} \label{subsec:cvae_construction_and_training}
\subsubsection{Data Preprocessing} \label{subsubsec:cvae_data_preprocessing}
The incident and reflected velocity data are preprocessed are scaled by $v_{\text{ref}}$ before being passed to the \gls{cvae} model.
To avoid numerical issues during training, the velocity components are scaled by a constant factor to bring them into a numerically convenient range.
The constant is chosen such that the scaled velocity magnitudes are approximately in the range $[0,\, 20]$.
Such a range is well suited to the activation functions employed in the network and avoids numerical issues such as exploding gradients.

\subsubsection{Hyperparameters} \label{subsubsec:cvae_hyperparameters}
The input for the encoder of the \gls{cvae} is the concatenation of the reflected velocity $\vecvr$ and the incident velocity $\vecvi$.
Both velocity vectors are three-dimensional, resulting in a six-dimensional input vector for the encoder.
The encoder consists of three hidden layers with 128, 64, and 32 neurons, respectively.
All hidden layers use the \gls{elu} activation function~\cite{clevert_elu_2016}.
For the latent space distribution a six-dimensional diagonal multivariate normal distribution is used.
This selection simplifies backpropagation during training and offers a closed form expression for the \gls{kl}-divergence term in the loss function, which is crucial for efficient training of the \gls{cvae}~\cite{kingma2022autoencodingvariationalbayes}.
Therefore, the encoder outputs a 12-dimensional vector, corresponding to the mean and the diagonal covariance of the latent normal distribution.
From the latent distribution, a six-dimensional latent vector $\bm{z}$ is sampled and concatenated with the incident velocity $\vecvi$ to form the input of the decoder.
As a result, the decoder has a nine-dimensional input vector.
For the decoder, three hidden layers with 32, 64, and 128 neurons are used.
The hidden layers of the decoder also use the \gls{elu} activation function.
In order to enforce the non-negativity constraint on the normal velocity component of the reflected velocity, the Softplus activation function is applied to the normal velocity component of the decoder's output.
The Softplus function avoids saturation issues that can arise with other activation functions, such as ReLU, while still ensuring that the output is strictly positive.
Here, the parameter $\beta$ of the Softplus function is set to $0.2$.
For the tangential velocity components, linear activation is used, allowing the model to learn the full range of possible tangential velocities without any constraints.
In comparison to the \gls{cvae} developed in~\citet{Schuette2025}, the architecture described here adds a hidden layer and increases the dimension of the latent space, allowing the model to capture more complex relationships in the data and learn a richer representation of the reflected velocity distribution from the thermal to hypersonic regime.

\subsubsection{Training} \label{subsubsec:cvae_training}
The loss function is twofold.
The first term is the \gls{mse} between the predicted reflected velocity $\vecvrp$ and the true reflected velocity $\vecvr$, encouraging the model to reconstruct the reflected velocity accurately.
The second term is the \gls{kl}-divergence between the encoder's predicted latent distribution and the standard normal prior $\mathcal{N}(0, \bm{I})$.
This term regularizes the latent space by penalizing deviations from the prior and promotes a smooth, well-structured representation suitable for sampling and generalization.
The \gls{kl}-divergence term is weighted by a factor of $0.8$ to balance the reconstruction accuracy and the regularization of the latent space.
For model training, the Adam optimizer is used for $80$ epochs and a batch size of $32$.
A step decay learning rate schedule is applied, where the learning rate is reduced by a factor of $0.1$ every $20$ epochs, starting from an initial value of $10^{-3}$.
This allows the model to make larger updates to the parameters in the early stages of training, which can help to escape local minima, and then make finer adjustments as training progresses to avoid overshooting the optimal parameters and to converge to a better solution.
The training of the model took approximately $10$ minutes.

% Model Construction and Training - cRealNVP ---------------------------------------------------- %
\subsection{\gls{crealnvp}} \label{subsec:crealnvp_construction_and_training}
\subsubsection{Data Preprocessing} \label{subsubsec:crealnvp_data_preprocessing}
For preprocessing the data for the \gls{crealnvp} model, the incident velocities $\vecvi$ and reflected $\vecvr$ are scaled using the same strategy as for the \gls{cvae}.
However, as already described in Section~\ref{subsubsec:crealnvp} the normal velocity component of the reflected velocity is constrained to be strictly positive.
To handle this constraint, the Softplus transformation is applied to the normal component after the initial factor scaling.
This transformation maps the positive normal velocity values to an unconstrained space, allowing the \gls{crealnvp} to operate in the full three-dimensional velocity space without violating the physical constraint on the normal velocity component.
For this, a $\beta$ parameter of $0.2$ is used in the Softplus function.
As discussed previously, this transformation does not need to be included in the loss function during training because the model learns the distribution in the transformed space.
However, when evaluating the model's log-probability in the physical velocity space, the Jacobian of the Softplus transformation must be included to get the correct probability density values.

\subsubsection{Hyperparameters} \label{subsubsec:crealnvp_hyperparameters}

The latent distribution in the \gls{crealnvp} is chosen to be a multivariate standard normal distribution $\mathcal{N}(0, \bm{I})$.
The dimension of the latent space is fixed by the dimension of the velocity space, resulting in a three-dimensional latent space.
In total $K=6$ coupling layers are used, allowing the model to learn complex transformations from the latent space to the velocity space.
Each coupling layer uses one neural network to predict the scaling $s$ and translating $t$ parameters of the affine transformation.
The neural network of each coupling layer consists of four hidden layers with 64 neurons each, using the \gls{relu} activation function~\cite{nair_relu_2010}.
Its output is split into two parts.
The first part corresponds to the scaling parameter $s$.
In order to ensure that the scaling factors are bounded, the output of this part is passed through the \gls{tanh} activation function.
This helps stabilize training and prevents extreme transformations that could lead to numerical issues.
The second part corresponds to the translating parameter $t$ and uses linear activation to allow unbounded translations.
The input to the neural network of each coupling layer is the concatenation of the preserved part ($x_{\text{on}}$ or $y_{\text{on}}$) and the incident velocity $\vecvi$.
For the coupling layers, a binary mask is used to determine which dimensions are transformed and which are preserved.
The mask configuration used across the six coupling layers is summarized in Table~\ref{tab:mask_pattern}.
This pattern ensures that all dimensions are transformed across the layers, allowing the model to capture complex interactions between the velocity components.

\begin{table}[tb!]
    \setlength{\tabcolsep}{10pt}
    \renewcommand{\arraystretch}{1.2}
    \centering
    \caption{Mask pattern for the coupling layers in the \gls{crealnvp} model for velocity components (t1, t2, n).}
    \begin{tabular}{l c c c}
        \toprule
        & \textbf{t1} & \textbf{t2} & \textbf{n} \\
        \midrule
        Layer 1 & off & on & off \\
        Layer 2 & off & off & on \\
        Layer 3 & on & off & off \\
        Layer 4 & on & on & off \\
        Layer 5 & off & on & on \\
        Layer 6 & on & off & on \\
        \bottomrule
    \end{tabular}
    \label{tab:mask_pattern}
\end{table}

\subsubsection{Training} \label{subsubsec:crealnvp_training}
Training of the \gls{crealnvp} model is performed by maximizing the exact conditional log-likelihood of the data.
This is equivalent to minimizing the \gls{nll} loss as defined in equation~\eqref{eq:nll_loss}.
Additionally, a detailed balance loss term, as defined in equation~\eqref{eq:loss_db}, is added to the training loss to encourage the model to satisfy the detailed balance requirement.
Therefore, physical constraints are incorporated into the training process.
In contrast, the \gls{cvae} does not offer a mechanism to impose detailed balance.
For evaluating the detailed balance loss, equilibrium data are used, consisting of incident and reflected velocity pairs sampled from a Maxwell flux distribution at $300\,\text{K}$.
This distribution is strongly concentrated at moderate velocity magnitudes around the thermal velocity.
As a consequence, the model receives detailed balance supervision primarily in this high probability region.
The loss resulting from low or high velocities contributes very little to the overall loss because samples in these tail regions of the distribution are sparse.
In practice, this means that good detailed balance performance is achieved only in the part of the velocity space where the equilibrium data are dense, whereas regions with low or large velocity magnitudes remain weakly constrained.
To increase the detailed balance performance for the tail regions, additional training samples are generated by drawing incident and reflected velocities uniformly between $0$ and $2\,000\,\text{m/s}$.
This augmentation increases the density of both low- and high magnitude velocities, enabling the model to learn the detailed balance relation across the entire equilibrium velocity space rather than only in the high probability region of the Maxwell flux distribution.
It is important to note that the specific distribution of incident and reflected velocities used in the detailed balance loss is irrelevant, since the detailed balance condition must hold for all velocity pairs in equilibrium.

At the start of the training, the detailed balance loss is weighted by a factor of $0.01$.
This allows the model to first learn the overall structure of the reflected velocity distribution from the data, without being overly constrained by the detailed balance requirement.
As training progresses, the weight of the detailed balance loss is increased by a factor of $10$ every $20$ epoch, gradually shifting more emphasis toward satisfying the detailed balance constraint once the model has learned the basic structure of the data distribution.
Training runs for $80$ epochs using the Adam optimizer and a batch size of $64$.
A step decay learning rate schedule is applied, reducing the learning rate by a factor of $0.1$ every $40$ epochs, starting from an initial value of $10^{-4}$.
This prevents degradation of the \gls{nll} performance while the detailed balance term becomes increasingly influential, which is essential to maintain an accurate fit to the hypersonic velocity distributions.
The training of the model took approximately $20$ minutes.

% ----------------------------------------------------------------------------------------------- %
% Physical Validation
% ----------------------------------------------------------------------------------------------- %
\section{Physical Validation} \label{sec:physical_validation}
Physical validation is carried out in two regimes: the equilibrium (Section~\ref{subsec:equilibrium}) and the non-equilibrium regime (Section~\ref{subsec:non_equilibrium}).
For equilibrium validation, the impacts of thermal velocities are examined.
The non equilibrium validation instead examines how well the model distributions reproduce the target distributions obtained from the \gls{md} simulations.

\subsection{Equilibrium} \label{subsec:equilibrium}
\begin{figure}[htbp]
    \centering
    \begin{subfigure}[t]{\columnwidth}
        \centering
        \includegraphics{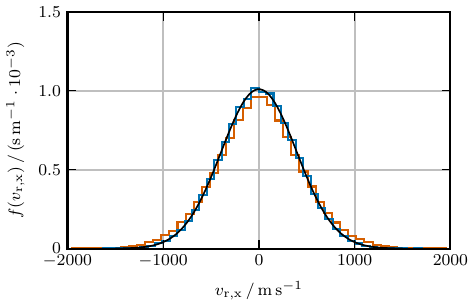}
        %\inputtikz{figures}{eq_flux_distribution_x}
    \end{subfigure}
    \begin{subfigure}[t]{\columnwidth}
        \centering
        \includegraphics{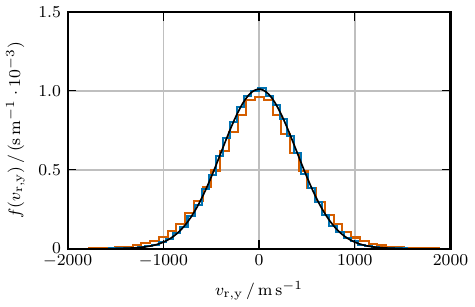}
        %\inputtikz{figures}{eq_flux_distribution_y}
    \end{subfigure}\hfill
    \begin{subfigure}[t]{\columnwidth}
        \centering
        \includegraphics{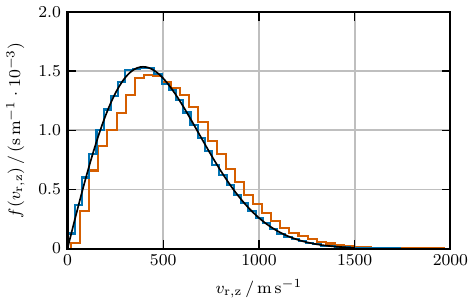}
        %\inputtikz{figures}{eq_flux_distribution_z}
    \end{subfigure}\hfill
    \begin{subfigure}[t]{\columnwidth}
        \centering
        \includegraphics{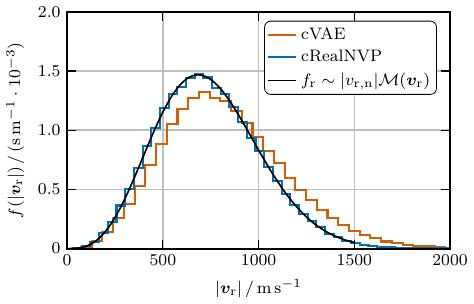}
        %\inputtikz{figures}{eq_flux_distribution_mag}
    \end{subfigure}
    \caption{Reflected velocity distributions of atomic oxygen for the \gls{cvae}, \gls{crealnvp}, and true reflected Maxwell flux distribution $\MaxwellFluxRe$ at the wall temperature $\Tw = 300\,\text{K}$, obtained for an incident Maxwell flux distribution $\MaxwellFluxIn$ at wall temperature.}

    \label{fig:histograms_eq_flux}
\end{figure}

If the gas and a surface are in thermal equilibrium, the reflected velocity distribution of an incident Maxwell flux distribution is also a Maxwell flux distribution with the same temperature.
This is a direct consequence of the detailed balance requirement, as already explained in Section~\ref{subsubsec:dsmc}.
In order to validate both models, $200\,000$ incident velocities are sampled from a Maxwell flux distribution at $300\,\text{K}$ and scattered at the surface using the \gls{cvae} and the \gls{crealnvp} models.
The reflected velocity distributions of the tangential and normal velocity components, as well as the velocity magnitude, of both models are then compared to the true reflected velocity distribution, which is a Maxwell flux distribution at $300\,\text{K}$.
This is shown in Figure~\ref{fig:histograms_eq_flux}.
While the \gls{crealnvp} shows very good agreement with the true reflected velocity distribution, the \gls{cvae} shows a worse performance, especially for the normal velocity component and the velocity magnitude.

\begin{figure}[t]
    \centering
    \begin{subfigure}{\columnwidth}
        \centering
        \includegraphics{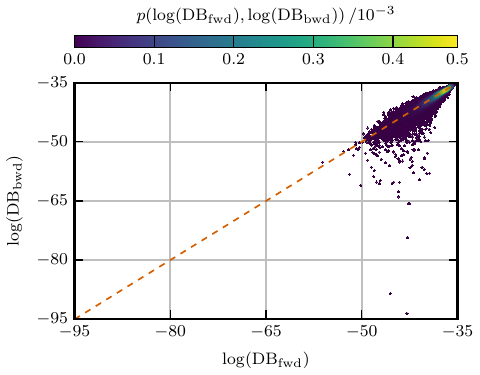}
        %\inputtikz{figures}{cRealNVP_no_db_loss_db_sides}
        \subcaption{\gls{crealnvp} without \gls{db}-loss}
        \label{subfig:db_sides_no_db_loss}
    \end{subfigure}
    \begin{subfigure}{\columnwidth}
        \centering
        \includegraphics{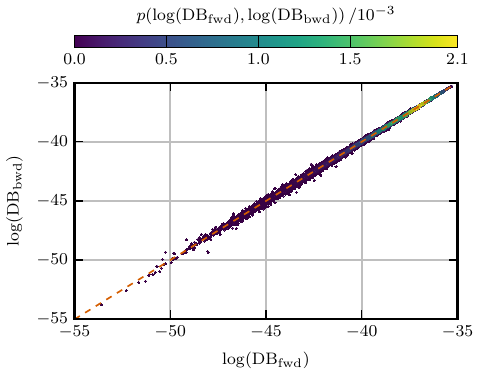}
        %\inputtikz{figures}{cRealNVP_db_sides}
        \subcaption{\gls{crealnvp} with \gls{db}-loss}
        \label{subfig:db_sides_db_loss}
    \end{subfigure}
    \caption{Forward and backward sides of the detailed balance (DB) relation for the scattering of an incident Maxwell flux distribution of atomic oxygen at $T = 300\,\text{K}$ at a surface with $\Tw = 300\,\text{K}$ for a \gls{crealnvp} model trained without and with a detailed balance loss term.}
    \label{fig:db_sides_plot}
\end{figure}

Compared to the \gls{cvae}, the \gls{crealnvp} has the advantage of directly evaluating the detailed balance relation, because it gives the exact conditional probability density function value for any incident and reflected velocity pair.
This allows to evaluate the detailed balance relation for the equilibrium data, which is shown in Figure~\ref{fig:db_sides_plot}.
In Figure~\ref{subfig:db_sides_db_loss} the detailed balance relation is evaluated for the \gls{crealnvp} model trained with the detailed balance loss.
To isolate the effect of this loss term, an otherwise identical \gls{crealnvp} model is trained without it.
Its detailed balance evaluation is shown in Figure~\ref{subfig:db_sides_no_db_loss}.

Exact satisfaction of the detailed balance requirement would produce identical forward (fwd) and backward (bwd) sides, causing all points to lie on the diagonal.
The model trained without the detailed balance loss exhibits substantial deviations from this diagonal, particularly at high velocity magnitudes.
These large discrepancies highlight the importance of including high magnitude samples in the detailed balance evaluation, as discussed earlier.

Incorporating the detailed balance loss into the training significantly reduces the discrepancy between the forward and backward sides, indicating that the model learns to approximate detailed balance with only small remaining deviations.
Nonetheless, the relation is not exactly satisfied and residual deviations remain.

% Detailed Balance in Flow-Based Scattering Kernels: Construction and Practical Limitations
A data-driven scattering kernel exactly satisfying detailed balance can be constructed analytically with a \gls{crealnvp} flow $p(\vecvr|\vecvi)$ by a geometric symmetrization.
The resulting kernel $\skernel{\vecvi}{\vecvr}$ is given by the following expression:
\begin{equation}
    \begin{aligned}
        \log \skernel{\vecvi}{\vecvr} = &\frac{1}{2} \left[ \log p(\vecvr|\vecvi) + \log p(-\vecvi|-\vecvr) \right] \\
        + &\frac{1}{2} \log \left( \frac{|\vecvr \cdot \bm{n}|\mathcal{M}(\vecvr) }{|\vecvi \cdot \bm{n}| \mathcal{M}(\vecvi) } \right)
    \end{aligned}
\end{equation}
By setting the forward $\skernel{\vecvi}{\vecvr}$ and the backward kernel $\skernel{-\vecvr}{-\vecvi}$ in relation, the flow term cancels exactly, yielding the detailed balance relation~\eqref{eq:db_req}
\begin{equation}
    \frac{\skernel{\vecvi}{\vecvr}}{\skernel{-\vecvr}{-\vecvi}} = \frac{|\vecvr \cdot \bm{n}|\mathcal{M}(\vecvr) }{|\vecvi \cdot \bm{n}| \mathcal{M}(\vecvi) }.
\end{equation}

Detailed balance is therefore satisfied by algebraic construction.
Despite its theoretical validity, this construction is not practically feasible, since detailed balance is now enforced even for non-equilibrium conditions.
The kernel above requires the evaluation of the backward flow $p(-\vecvi|-\vecvr)$, which represents the probability that a particle arriving at the wall with velocity $-\vecvr$ is reflected with velocity $-\vecvi$.
In non-equilibrium conditions, such reverse events are thermodynamically suppressed.
A particle incident at $8\,000\,\text{m/s}$ loses energy during the scattering process with high probability, resulting in lower velocities such as $5\,000\,\text{m/s}$.
The reverse event, in which a particle incident at $5\,000\,\text{m/s}$ gains energy to reach $8\,000\,\text{m/s}$, is highly unlikely and is essentially absent from the \gls{md} data.
Therefore, the backward inputs $(-\vecvi, -\vecvr)$ lie arbitrarily far outside the training distribution of the \gls{crealnvp} model, making the evaluation of $p(-\vecvi|-\vecvr)$ numerically meaningless in the non-equilibrium regime.

A natural solution would be to construct artificial backward data by mirroring the forward \gls{md} dataset ${(\vecvr|\vecvi)}_{\text{\gls{md}}}$ to ${(-\vecvi|-\vecvr)}_{\text{mirrored}}$.
However, this does not resolve the physical inconsistency.
The mirrored data implicitly claim that the probability of a transition from $8\,000\,\text{m/s}$ to $5\,000\,\text{m/s}$ equals the probability of the reverse transition from $5\,000\,\text{m/s}$ to $8\,000\,\text{m/s}$, which contradicts the thermodynamics of the non-equilibrium process.

A detailed balance satisfying kernel constructed from \gls{md} data is only reliable near equilibrium, where reverse events occur with appreciable probability and the backward flow can be evaluated within the training distribution.
Far from equilibrium, the backward path is physically suppressed and numerically inaccessible.
Forward and backward rates are no longer balanced, and enforcing detailed balance would move the model away from the scattering statistics observed in the \gls{md} data.
Therefore, the hybrid strategy adopted in this work provides an appropriate balance between physical fidelity and data driven flexibility.
Near equilibrium, where detailed balance is a fundamental physical requirement, a dedicated loss term steers the model toward approximate satisfaction of this condition.
In non-equilibrium settings, the kernel is instead trained to reproduce the scattering statistics observed in the \gls{md} data.
This formulation yields the best achievable behavior across both regimes: approximate detailed balance where it is physically meaningful, and accurate reproduction of non-equilibrium scattering statistics.

The Appendix~\ref{secapdix:eq_validation} additionally presents scattering results under equilibrium conditions for different wall temperatures $\Tw$ and species.
As shown there, the \gls{crealnvp} model accurately reproduces the correct distributions across all considered cases.
This generalization capability stems directly from the scaling approach introduced in Section~\ref{subsec:velocity_scaling}.
By scaling the velocities with the thermal velocity $v_{\text{th}}$, the model learns the underlying distributions in a normalized space that is independent of wall temperature and species.
When the model is applied to different wall temperatures or species, the velocities are scaled using the corresponding thermal velocity $v_{\text{th}}$ for that specific condition.
This allows the model to adapt to different wall temperatures and species without requiring additional training data, as the learned distributions are inherently generalizable across these conditions.
As mentioned in Section~\ref{subsec:velocity_scaling}, while the species scaling works well in equilibrium, it is not expected to generalize across species in non-equilibrium, since the interaction potential in \gls{md} simulations differs for each species.

\subsection{Non-Equilibrium} \label{subsec:non_equilibrium}
For the non-equilibrium validation, the reflected velocity distribution of incident velocity magnitudes of $3\,000\,\text{m/s}$, $7\,261.3\,\text{m/s}$, and $8585.9\,\text{m/s}$ are compared to the true reflected velocity distribution obtained from the \gls{md} simulations.
The models haven't seen any data for these three velocity magnitudes during training, so this validation tests the interpolation capabilities of the models.
The distributions of the tangential and normal velocity components for the \gls{cvae} and the \gls{crealnvp} models are shown in Figures~\ref{fig:histograms_v_mag_3000}, \ref{fig:histograms_v_mag_7261.3}, and \ref{fig:histograms_v_mag_8585.9} for incident polar angles of $0^\circ$, $20^\circ$, $40^\circ$, $60^\circ$, and $80^\circ$.
In general, both models show a good agreement with the true reflected velocity distribution in all directions for all incident velocity magnitudes and incident polar angles.
However, the \gls{crealnvp} model shows better agreement with the true reflected velocity distribution, especially for the normal velocity component and the $t1$-direction at high incident polar angles.
Nevertheless, the \gls{cvae} model also captures the overall shape of the reflected velocity distribution, but struggles to capture the peak in the $t1$-direction at an incident polar angle of $80^\circ$ as can be seen in Figures~\ref{subfig:histogram_v_mag_8585.9_80} and \ref{subfig:histogram_v_mag_7261.3_80}.

\begin{figure}[tbp]
    \centering
    \begin{subfigure}[t]{\columnwidth}
        \centering
        \includegraphics{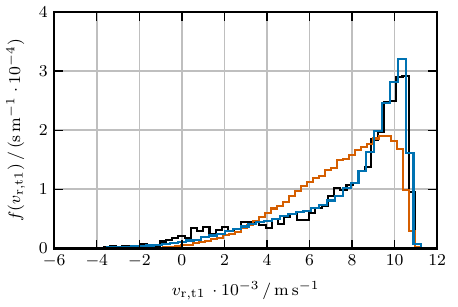}
        %\inputtikz{figures}{reflected_velocity_distribution_t1_11000_80}
    \end{subfigure}
    \begin{subfigure}[t]{\columnwidth}
        \centering
        \includegraphics{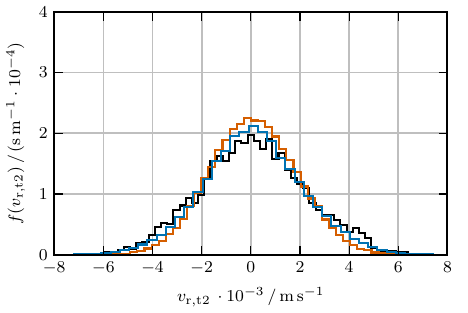}
        %\inputtikz{figures}{reflected_velocity_distribution_t2_11000_80}
    \end{subfigure}
    \begin{subfigure}[t]{\columnwidth}
        \centering
        \includegraphics{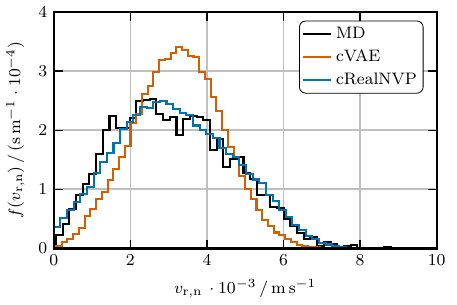}
        %\inputtikz{figures}{reflected_velocity_distribution_n_11000_80}
    \end{subfigure}
    \caption{Reflected velocity distributions of atomic oxygen for the \gls{cvae}, \gls{crealnvp} and \gls{md} data for an incident velocity magnitude of $11\,000\,\text{m/s}$ and incident polar angle of $80^\circ$ at a wall temperature of $\Tw = 300\,\text{K}$.}
    \label{fig:histrograms_v_mag_11000_80}
\end{figure}

In addition to the validation data, \gls{md} simulations are performed for an incident velocity magnitude of $11\,000\,\text{m/s}$ and an incident polar angle of $80^\circ$.
This velocity magnitude is way outside the velocity range that occurs under \gls{vleo} conditions and the reflected velocity distribution is used to evaluate the extrapolation capabilities of the models.
In Figure~\ref{fig:histrograms_v_mag_11000_80} the reflected velocity distribution of the \gls{cvae} and the \gls{crealnvp} models are compared to the true reflected velocity distribution.
As can be seen, the \gls{crealnvp} shows tremendous extrapolation capabilities, with very good agreement with the true reflected velocity distribution.
On the other hand, the \gls{cvae} model struggles to capture the reflected velocity distribution, especially for the normal velocity component and the $t1$-direction.
Even attempts with deeper architectures and larger latent space dimensions fail to improve the performance of the \gls{cvae}, despite their potential to capture more complex relationships in the data.

As a result, the \gls{crealnvp} model has the best accuracy for the non-equilibrium and the equilibrium regime.
The \gls{cvae} still achieves an acceptable accuracy in the non-equilibrium regime, but struggles on the extrapolation and the equilibrium regime.

In Appendix~\ref{secapdix:neq_validation_wall_temp}, additional non-equilibrium validation results are presented for different wall temperatures $\Tw$.
Good agreement with the reference \gls{md} data is observed for all considered wall temperatures.
This indicates that the scaling approach also seems to be effective in the non-equilibrium regime, allowing the models to generalize across different wall temperatures without requiring additional training data.
However, it should be noted that this scaling still requires validation across a broader range of conditions.
Ideally, future work would include non-equilibrium impact data at several wall temperatures directly in the training process.
This would allow the model to learn the wall temperature dependence directly from data.
However, generating \gls{md} data for multiple wall temperatures is computationally expensive.
This makes the proposed scaling approach a reasonable and pragmatic solution at this stage.

% ----------------------------------------------------------------------------------------------- %
% Application
% ----------------------------------------------------------------------------------------------- %
\section{Application} \label{sec:application}

Two applications of the developed scattering kernels are presented in this Section.
Section~\ref{subsec:reservoir_simulation} demonstrates their use in a \gls{dsmc} reservoir simulation, illustrating the models' behavior in (near-) equilibrium.
Section~\ref{subsec:aerodynamic_coefficients} applies the kernels in a \gls{dsmc} simulation for computing aerodynamic coefficients of a flat plate under \gls{vleo} conditions, thereby assessing their non-equilibrium performance.

For both applications, the scattering kernels are implemented in a \gls{dsmc} framework.
In this context, the \gls{dsmc} method implemented in PICLas can be applied~\cite{fasoulas_combining_2019, piclas_github}.
PICLas is an open-source particle-based simulation code for rarefied gas flows, developed at the Institute of Space Systems at the University of Stuttgart and the company boltzplatz.

% Application - Reservoir Simulation
\subsection{Reservoir Simulation} \label{subsec:reservoir_simulation}

\begin{figure}[tbp]
    \centering
    \includegraphics{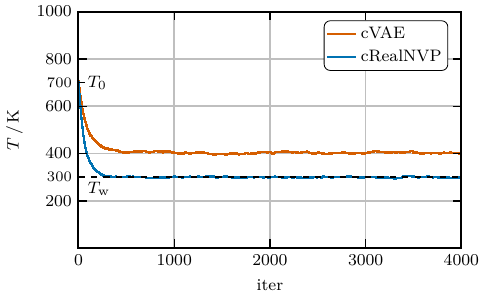}
    %\inputtikz{figures}{reservoir_gas_temperature}
    \caption{Evolution of the gas temperature in the reservoir simulation for atomic oxygen, starting from an initial temperature of $T_0=700\,\text{K}$ and a wall temperature of $T_{\text{w}}=300\,\text{K}$ for the \gls{cvae} and the \gls{crealnvp} models.}
    \label{fig:reservoir_gas_temperature}
\end{figure}

For reservoir simulation, a simple setup is used, where a gas is confined in a cubic box with the size of $1\,\text{m}$.
The gas is initialized with $10\,000$ particles, where the velocity of each particle is sampled from a Maxwell distribution at a temperature of $T_0=700\,\text{K}$.
The wall temperature is set to $T_{\text{w}}=300\,\text{K}$.
Since collisions between gas particles are neglected, the only mechanism for energy exchange is gas-surface scattering at the walls.
A time step of $10^{-5}\,\text{s}$ is used for the simulation, and the number of iterations is set to $4\,000$.
Both scattering kernels \gls{cvae} and \gls{crealnvp} are used to model the scattering on the walls, and the resulting gas temperature in the reservoir is monitored over time.
The progress of the gas temperature in the reservoir is shown in Figure~\ref{fig:reservoir_gas_temperature}.
For the \gls{crealnvp} model, the gas temperature converges to the wall temperature.
The mean gas temperature of the last $3\,000$ iterations is approximately $300.29\,\text{K}$.
In contrast, the \gls{cvae} model struggles to reach the wall temperature.
Its mean gas temperature reaches approximately $401.96\,\text{K}$, far above the prescribed value of $300\,\text{K}$.
This behavior is consistent with the results in Section~\ref{subsec:equilibrium}, where the \gls{cvae} model already struggled to reproduce the reflected velocity distribution for an incident Maxwell flux distribution at $300\,\text{K}$ of atomic oxygen.

Appendix~\ref{secapdix:reservoir_application} presents additional reservoir simulation results for different wall temperatures $\Tw$ and species.
These results show that the \gls{crealnvp} model accurately reproduces the correct wall temperature in all considered cases, confirming that the scaling approach works correctly in the equilibrium regime.

% Application - Aerodynamic Coefficients
\subsection{Aerodynamic Coefficients} \label{subsec:aerodynamic_coefficients}

\begin{figure}[tb!]
    \centering
    \includegraphics{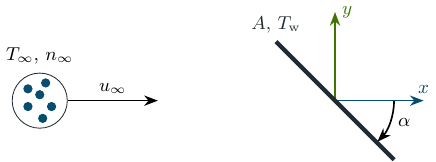}
    %\inputtikz{figures}{sketch_aerodynamic_coefficients}
    \caption{Schematic representation of the \gls{dsmc} simulation setup used to compute aerodynamic coefficients.}
    \label{fig:aerodynamic_coefficients}
\end{figure}

\begin{figure}[tb!]
    \centering
    \begin{subfigure}[t]{\columnwidth}
        \centering
        \includegraphics{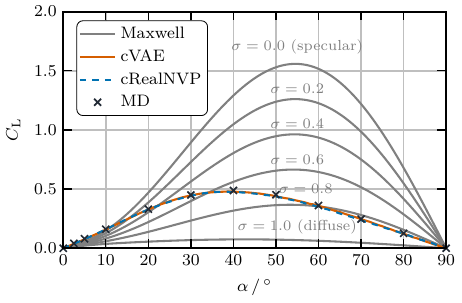}
        %\inputtikz{figures}{lift_coefficient}
        \caption{Lift Coefficient}
    \end{subfigure}
    \begin{subfigure}[t]{\columnwidth}
        \centering
        \includegraphics{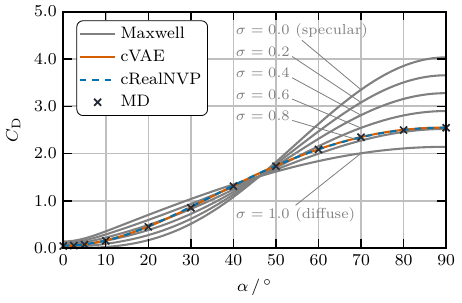}
        %\inputtikz{figures}{drag_coefficient}
        \caption{Drag Coefficient}
    \end{subfigure}
    \begin{subfigure}[t]{\columnwidth}
        \centering
        \includegraphics{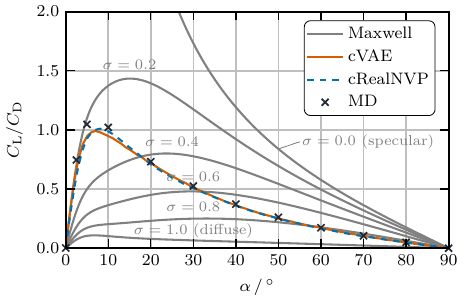}
        %\inputtikz{figures}{lift_to_drag_ratio}
        \caption{Lift-to-Drag Ratio}
    \end{subfigure}
    \caption{Comparison of aerodynamic coefficients for atomic oxygen between the \gls{cvae}, \gls{crealnvp}, \gls{md} reflection data, and Maxwell model (with $T_{\text{r}} = T_{\text{w}} = 300\,\text{K}$) for a flat plate in \gls{vleo} conditions.}
    \label{fig:aerodynamiccoefficients}
\end{figure}

For the application of the scattering kernels in a non-equilibrium scenario, a simple flat plate is considered to compute aerodynamic coefficients under \gls{vleo} conditions.
The \gls{dsmc} simulation setup is shown in Figure~\ref{fig:aerodynamic_coefficients}.
The flat plate has an area of $A=0.09\,\text{m}^2$ and a wall temperature of $\Tw = 300\,\text{K}$.
As incident conditions, a bulk velocity magnitude of $u_{\infty}=7\,800\,\text{m/s}$, a temperature of $T_{\infty} = 934\,\text{K}$, and an atomic oxygen number density of $n_{\text{O}, \infty} = 4.698\cdot 10^{14}\,\text{m}^{-3}$ are used, which correspond to typical \gls{vleo} conditions.
The angle of attack is varied between $0^\circ$ and $90^\circ$ in steps of $1^\circ$.
In every simulation, $100\,000$ particle surface interactions are simulated and the resulting aerodynamic coefficients are calculated.
The aerodynamic coefficients are derived from the total force $\bm{F}$ acting on the flat plate, which is calculated from the momentum exchange of the particles with the surface.
The lift coefficient $C_{\text{L}}$ and the drag coefficient $C_{\text{D}}$ are calculated using the following formulas:
\begin{equation}
    C_{\text{L}} = \frac{F_{\text{y}}}{\frac{1}{2} m_{\text{O}} n_{\text{O},\infty} u_{\infty}^2 A}
\end{equation}
\begin{equation}
    C_{\text{D}} = \frac{F_{\text{x}}}{\frac{1}{2} m_{\text{O}} n_{\text{O},\infty} u_{\infty}^2 A}
\end{equation}
where $m_{\text{O}}=2.657 \cdot 10^{-26}\,\text{kg}$ is the mass of an atomic oxygen particle.
In Figure~\ref{fig:aerodynamiccoefficients} the lift coefficient, drag coefficient and lift-to-drag ratio are shown for the \gls{cvae} and the \gls{crealnvp} models.
Additionally, the aerodynamic coefficients of the Maxwell model with $T_{\text{r}} = T_{\text{w}}$ are shown for comparison for different accommodation coefficients $\sigma$, which is currently the most common model for satellite aerodynamics in \gls{vleo}.
These coefficients can be analytically calculated for a flat plate using the formulas derived by Bird~\cite{bird_molecular_1994}.

Furthermore, for selected angles of attack, aerodynamic coefficients are calculated approximately from the \gls{md} data, as a reference for the \gls{cvae} and the \gls{crealnvp} models.
For this purpose, the mean momentum exchange of the reflected velocities is calculated with respect to each selected incident velocity.
To calculate the forces acting on the flat plate, the incident velocity distribution is divided into bins of velocity magnitude and the incident polar angle.
For each bin, we assign the previously calculated mean momentum exchange for the corresponding incident velocity.
Afterwards, the total force is calculated by summing over all bins, where the contribution of each bin is weighted by the number of particles in that bin.
It is important to note that this approach is only an approximation, since the resolution of the bins is highly limited by the amount of \gls{md} data available.
Experimental data suitable for validating the models would be highly valuable, but there are no measurements for the specific \gls{vleo} conditions and surface material considered in this work.

The comparison of the aerodynamic coefficients of the \gls{cvae} and the \gls{crealnvp} models to the \gls{md} reference data in Figure~\ref{fig:aerodynamiccoefficients} shows that both models are capable of reproducing the \gls{md} reference data well, while the Maxwell model significantly deviates.
However, the \gls{crealnvp} shows slightly better agreement in the lift-to-drag ratio at low angles of attack compared to the \gls{cvae} model.
This is a result of the better performance of the \gls{crealnvp} model for the reflected velocity distribution at high incident polar angles (with respect to the surface normal) as discussed in Section~\ref{subsec:non_equilibrium}.
On the other hand, the \gls{cvae} allows for significantly faster calculations, as the sampling of reflected velocities is much faster compared to the \gls{crealnvp} model.
For $100\,000$ particle surface interactions, the \gls{cvae} is approximately $45\,\%$ faster than the \gls{crealnvp} model.
This results from the fact that the decoder of the \gls{cvae} consists of only three hidden layers, while the \gls{crealnvp} model consists of six coupling layers, where each coupling layer has a neural network with four hidden layers.
Therefore, if computational performance is of high importance and accuracy in the equilibrium regime is not crucial, the \gls{cvae} model can be a good choice for the scattering kernel.

In summary, the \gls{crealnvp} shows impressive performance in the equilibrium and non-equilibrium regime, making it applicable for a wide range of application scenarios going from thermal to hypersonic velocities.

% ----------------------------------------------------------------------------------------------- %
% Conclusion
% ----------------------------------------------------------------------------------------------- %
\section{Conclusion} \label{sec:conclusion}

In this work, we develop machine learning-based scattering kernels capable of bridging thermal to hypersonic velocity regimes.
The previously developed machine learning model was limited to hypersonic velocities that occur in \gls{vleo} and single scattering events.
We extended this by augmenting the \gls{md} training data with low-velocity impact simulations and equilibrium data.
This allows the model to be used across a wide range of application scenarios, from thermal to hypersonic velocities.
Especially enabling multi-scattering events, which is crucial for \gls{abep} intake development of \gls{vleo} satellites and surface roughness investigations.

We trained and compared two generative models: a \gls{cvae} and a \gls{crealnvp} flow model.
The \gls{crealnvp} architecture permits direct evaluation of the scattering kernel probability density $\skernel{\vecvi}{\vecvr}$.
This model demonstrated superior performance in the equilibrium and non-equilibrium regimes.
It accurately reproduced the reflected velocity distributions, approximately satisfies detailed balance constraints, and achieved proper thermal convergence in reservoir simulations.
In contrast, the \gls{cvae} model exhibited deficiencies in these areas.

In the non-equilibrium regime, both models showed good agreement with \gls{md} reference data.
However, the \gls{crealnvp} model maintained better accuracy for high incident polar angles.
It also demonstrated robust extrapolation to velocity magnitudes beyond the training range.
The application to aerodynamic coefficient calculations for a flat plate under \gls{vleo} conditions confirmed that both models reproduce \gls{md} predictions well.
On the other hand, the \gls{cvae} offers substantially better computational performance.
Therefore, it may be suitable for applications where equilibrium accuracy is less critical and computational efficiency is of importance.

To generalize the models across different wall temperatures and species, a sigmoid-based scaling approach was introduced.
This scaling showed good agreement with \gls{md} reference data for different wall temperatures in non-equilibrium, and was confirmed by the theoretical behavior in equilibrium.
However, this represents a first approach and requires further validation across a broader range of conditions.
Ideally, future work would include non-equilibrium training data at multiple wall temperatures, with $\Tw$ as an explicit conditioning variable.
Ultimately, our goal is to incorporate chemical reactions into the gas-surface interaction framework to achieve a complete model for reactive flow simulations in \gls{vleo} environments.

% ----------------------------------------------------------------------------------------------- %
% Acknowledgments
% ----------------------------------------------------------------------------------------------- %
\section*{Acknowledgments}
This work is funded by the Deutsche Forschungsgemeinschaft project number 516238647 - SFB1667/1 (ATLAS - Advancing Technologies for Low-Altitude Satellites). The authors also thank the High
Performance Computing Center Stuttgart (HLRS) for granting the computational time that allowed the execution of the presented molecular dynamics simulations.

% ----------------------------------------------------------------------------------------------- %
% Data Availability Statement
% ----------------------------------------------------------------------------------------------- %
\section*{Data Availability Statement}
Data and code supporting the findings of this study are available from the corresponding author upon reasonable request.

% ----------------------------------------------------------------------------------------------- %
% Conflict of Interest
% ----------------------------------------------------------------------------------------------- %
\section*{Conflict of Interest}
The authors declare that they have no conflicts of interest, financial or otherwise, that could have influenced the results or interpretation of the reported research.

% ----------------------------------------------------------------------------------------------- %
% Bibliography
% ----------------------------------------------------------------------------------------------- %
%\bibliographystyle{}
\bibliography{library}

% ----------------------------------------------------------------------------------------------- %
% Appendix
% ----------------------------------------------------------------------------------------------- %
\appendix
\onecolumngrid

% ----------------------------------------------------------------------------------------------- %
% Appendix - Validation
% ----------------------------------------------------------------------------------------------- %
\newpage

\section{Equilibrium Validation Across Wall Temperatures and Species} \label{secapdix:eq_validation}
\vspace{-8mm}
\begin{figure}[H]
    \centering
    \begin{minipage}{\textwidth}
        \centering
        \begin{minipage}[b]{0.32\textwidth}
            \centering
            \includegraphics{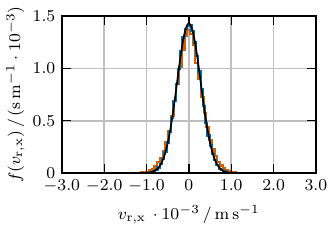}
            %\inputtikz{figures}{eq_flux_distribution_x_O2}
        \end{minipage}\hfill%
        \begin{minipage}[b]{0.32\textwidth}
            \centering
            \includegraphics{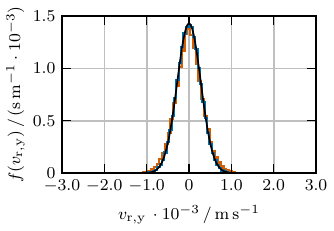}
            %\inputtikz{figures}{eq_flux_distribution_y_O2}
        \end{minipage}\hfill%
        \begin{minipage}[b]{0.32\textwidth}
            \centering
            \includegraphics{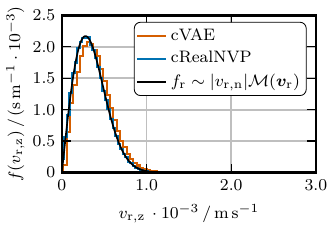}
            %\inputtikz{figures}{eq_flux_distribution_z_O2}
        \end{minipage}
        \subcaption{Molecular oxygen and $\Tw = 300\,\text{K}$}
    \end{minipage}
    \begin{minipage}{\textwidth}
        \centering
        \begin{minipage}[b]{0.32\textwidth}
            \centering
            \includegraphics{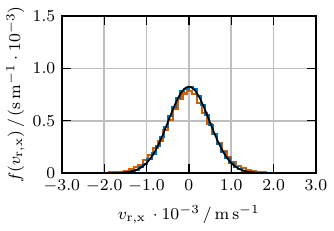}
            %\inputtikz{figures}{eq_flux_distribution_x_O2_900}
        \end{minipage}\hfill%
        \begin{minipage}[b]{0.32\textwidth}
            \centering
            \includegraphics{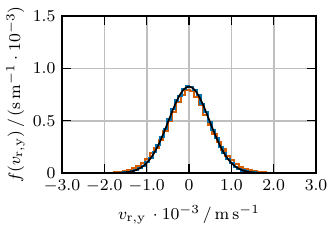}
            %\inputtikz{figures}{eq_flux_distribution_y_O2_900}
        \end{minipage}\hfill%
        \begin{minipage}[b]{0.32\textwidth}
            \centering
            \includegraphics{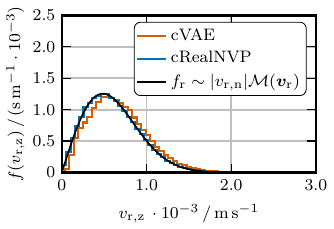}
            %\inputtikz{figures}{eq_flux_distribution_z_O2_900}
        \end{minipage}
        \subcaption{Molecular oxygen and $\Tw = 900\,\text{K}$}
    \end{minipage}
    \begin{minipage}{\textwidth}
        \centering
        \begin{minipage}[b]{0.32\textwidth}
            \centering
            \includegraphics{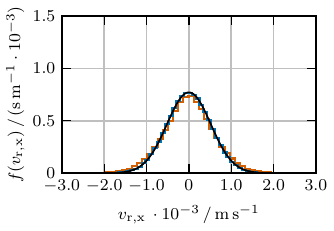}
            %\inputtikz{figures}{eq_flux_distribution_x_N2_900}
        \end{minipage}\hfill%
        \begin{minipage}[b]{0.32\textwidth}
            \centering
            \includegraphics{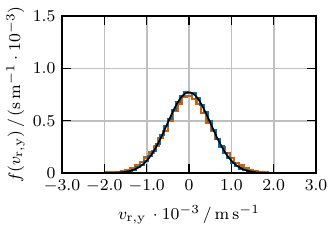}
            %\inputtikz{figures}{eq_flux_distribution_y_N2_900}
        \end{minipage}\hfill%
        \begin{minipage}[b]{0.32\textwidth}
            \centering
            \includegraphics{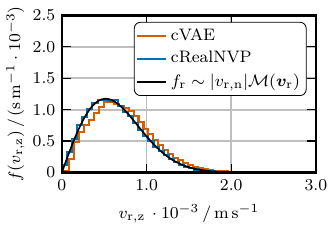}
            %\inputtikz{figures}{eq_flux_distribution_z_N2_900}
        \end{minipage}
        \subcaption{Molecular nitrogen and $\Tw = 900\,\text{K}$}
    \end{minipage}
    \begin{minipage}{\textwidth}
        \centering
        \begin{minipage}[b]{0.32\textwidth}
            \centering
            \includegraphics{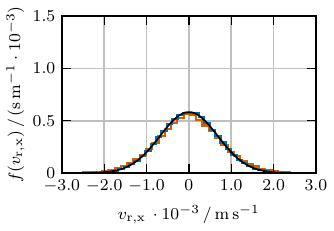}
            %\inputtikz{figures}{eq_flux_distribution_x_O_900}
        \end{minipage}\hfill%
        \begin{minipage}[b]{0.32\textwidth}
            \centering
            \includegraphics{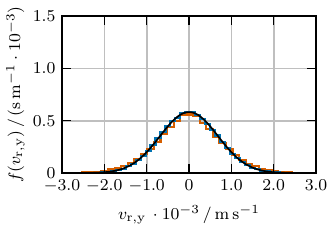}
            %\inputtikz{figures}{eq_flux_distribution_y_O_900}
        \end{minipage}\hfill%
        \begin{minipage}[b]{0.32\textwidth}
            \centering
            \includegraphics{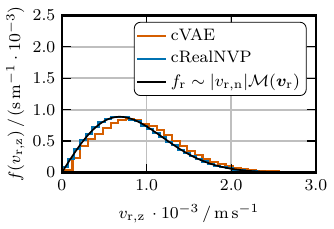}
            %\inputtikz{figures}{eq_flux_distribution_z_O_900}
        \end{minipage}
        \subcaption{Atomic oxygen and $\Tw = 900\,\text{K}$}
    \end{minipage}
    \begin{minipage}{\textwidth}
        \centering
        \begin{minipage}[b]{0.32\textwidth}
            \centering
            \includegraphics{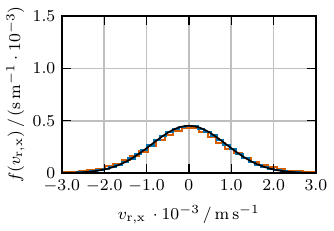}
            %\inputtikz{figures}{eq_flux_distribution_x_O_1500}
        \end{minipage}\hfill%
        \begin{minipage}[b]{0.32\textwidth}
            \centering
            \includegraphics{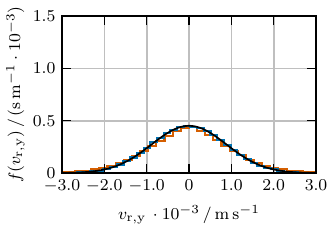}
            %\inputtikz{figures}{eq_flux_distribution_y_O_1500}
        \end{minipage}\hfill%
        \begin{minipage}[b]{0.32\textwidth}
            \centering
            \includegraphics{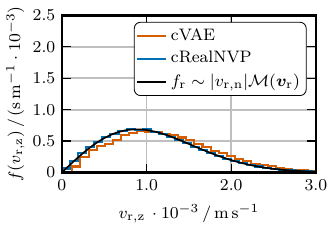}
            %\inputtikz{figures}{eq_flux_distribution_z_O_1500}
        \end{minipage}
        \subcaption{Atomic oxygen and $\Tw = 1500\,\text{K}$}
    \end{minipage}
    \caption{Reflected velocity distributions for the \gls{cvae}, \gls{crealnvp}, and true reflected Maxwell flux distribution $\MaxwellFluxRe$ for different species and wall temperatures, obtained for an incident Maxwell flux distribution $\MaxwellFluxIn$ at wall temperature.}
    \label{fig:histograms_eq_flux_appx}
\end{figure}

\begin{figure}[H]
    \centering
    \subcaptionbox{Molecular oxygen and $\Tw = 300\,\text{K}$}[0.45\textwidth]{%
        \includegraphics{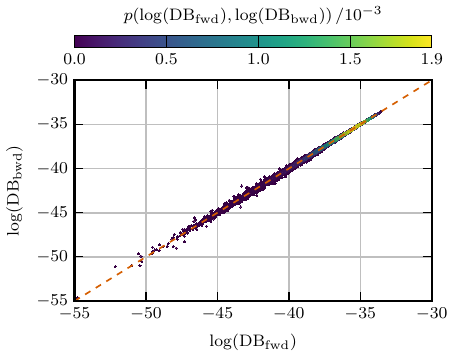}%
        %\inputtikz{figures}{cRealNVP_db_sides_O2}%
    }%
    \hfill%
    \subcaptionbox{Molecular oxygen and $\Tw = 900\,\text{K}$}[0.45\textwidth]{%
        \includegraphics{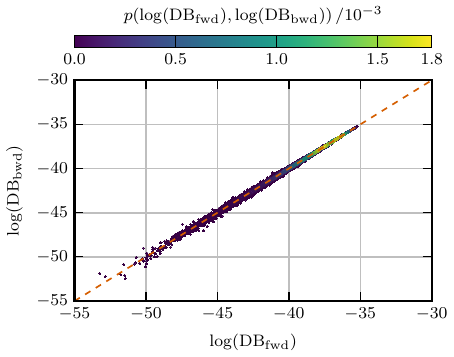}%
        %\inputtikz{figures}{cRealNVP_db_sides_O2_900}%
    }%
    \\[6pt]
    \subcaptionbox{Molecular nitrogen and $\Tw = 900\,\text{K}$}[0.45\textwidth]{%
        \includegraphics{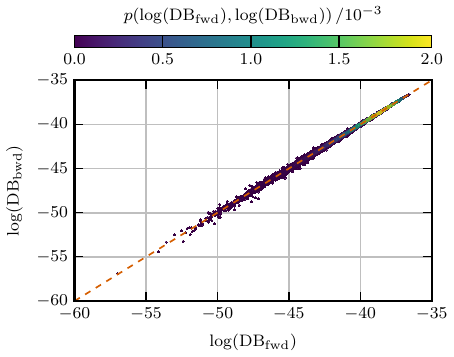}%
        %\inputtikz{figures}{cRealNVP_db_sides_N2_900}%
    }%
    \hfill%
    \subcaptionbox{Atomic oxygen and $\Tw = 900\,\text{K}$}[0.45\textwidth]{%
        \includegraphics{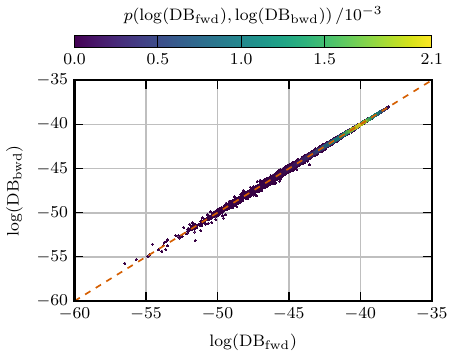}%
        %\inputtikz{figures}{cRealNVP_db_sides_O_900}%
    }%
    \\[6pt]
    \subcaptionbox{Atomic oxygen and $\Tw = 1500\,\text{K}$}[0.45\textwidth]{%
        \includegraphics{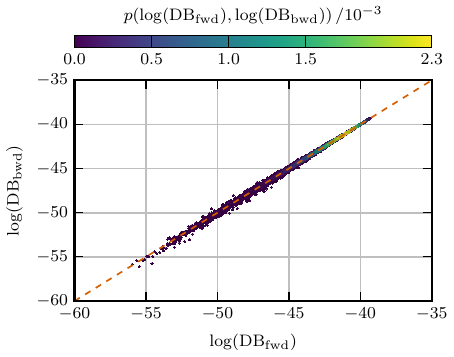}%
        %\inputtikz{figures}{cRealNVP_db_sides_O_1500}%
    }%
    \hfill%
    \subcaptionbox{Molecular oxygen and $\Tw = 1500\,\text{K}$}[0.45\textwidth]{%
        \includegraphics{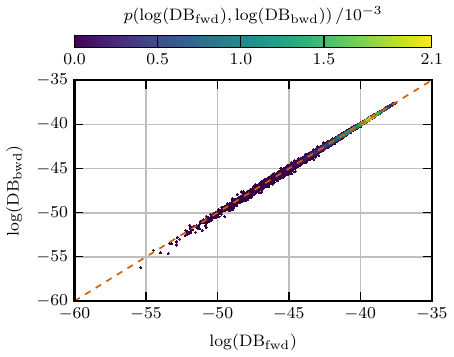}%
        %\inputtikz{figures}{cRealNVP_db_sides_O2_1500}%
    }%
    \vspace{-2mm}
    \caption{Forward and backward sides of the detailed balance (DB) relation for the scattering of an incident Maxwell flux distribution at wall temperature, evaluated for different species and wall temperatures using the \gls{crealnvp} model trained with a detailed balance loss term.}
    \label{fig:db_sides_plot_appx}
\end{figure}

\section{Non-Equilibrium Validation} \label{secapdix:neq_validation}
\vspace{-8mm}
\begin{figure}[H]
    \centering
    \begin{minipage}{\textwidth}
        \centering
        \begin{minipage}[b]{0.32\textwidth}
            \centering
            \includegraphics{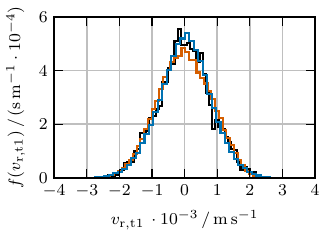}
            %\inputtikz{figures}{reflected_velocity_distribution_t1_3000.0_0}
        \end{minipage}\hfill%
        \begin{minipage}[b]{0.32\textwidth}
            \centering
            \includegraphics{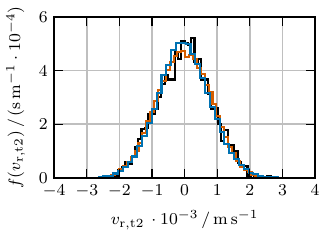}
            %\inputtikz{figures}{reflected_velocity_distribution_t2_3000.0_0}
        \end{minipage}\hfill%
        \begin{minipage}[b]{0.32\textwidth}
            \centering
            \includegraphics{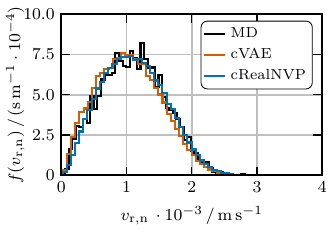}
            %\inputtikz{figures}{reflected_velocity_distribution_n_3000.0_0}
        \end{minipage}
        \vspace{-2mm}
        \subcaption{$|\bm{v}_{\text{i}}| = 3000.0 \, \text{m}/\text{s}$, $\theta_{\text{i}} = 0.0 \, ^\circ$}
    \end{minipage}
    \begin{minipage}{\textwidth}
        \centering
        \begin{minipage}[b]{0.32\textwidth}
            \centering
            \includegraphics{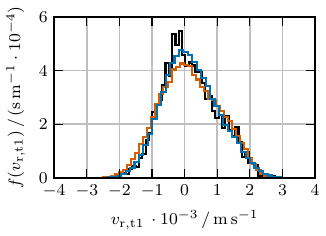}
            %\inputtikz{figures}{reflected_velocity_distribution_t1_3000.0_20}
        \end{minipage}\hfill%
        \begin{minipage}[b]{0.32\textwidth}
            \centering
            \includegraphics{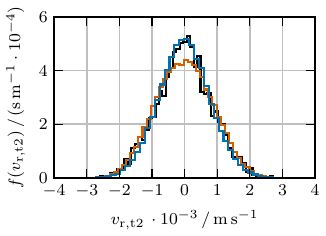}
            %\inputtikz{figures}{reflected_velocity_distribution_t2_3000.0_20}
        \end{minipage}\hfill%
        \begin{minipage}[b]{0.32\textwidth}
            \centering
            \includegraphics{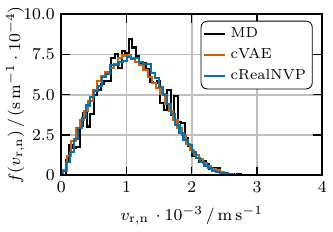}
            %\inputtikz{figures}{reflected_velocity_distribution_n_3000.0_20}
        \end{minipage}
        \vspace{-2mm}
        \subcaption{$|\bm{v}_{\text{i}}| = 3000.0 \, \text{m}/\text{s}$, $\theta_{\text{i}} = 20.0 \, ^\circ$}
    \end{minipage}
    \begin{minipage}{\textwidth}
        \centering
        \begin{minipage}[b]{0.32\textwidth}
            \centering
            \includegraphics{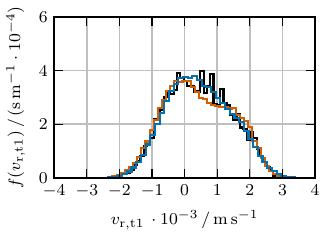}
            %\inputtikz{figures}{reflected_velocity_distribution_t1_3000.0_40}
        \end{minipage}\hfill%
        \begin{minipage}[b]{0.32\textwidth}
            \centering
            \includegraphics{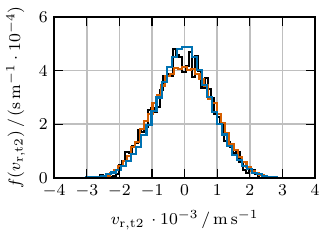}
            %\inputtikz{figures}{reflected_velocity_distribution_t2_3000.0_40}
        \end{minipage}\hfill%
        \begin{minipage}[b]{0.32\textwidth}
            \centering
            \includegraphics{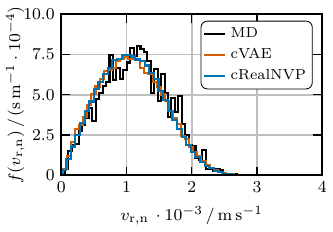}
            %\inputtikz{figures}{reflected_velocity_distribution_n_3000.0_40}
        \end{minipage}
        \vspace{-2mm}
        \subcaption{$|\bm{v}_{\text{i}}| = 3000.0 \, \text{m}/\text{s}$, $\theta_{\text{i}} = 40.0 \, ^\circ$}
    \end{minipage}
    \begin{minipage}{\textwidth}
        \centering
        \begin{minipage}[b]{0.32\textwidth}
            \centering
            \includegraphics{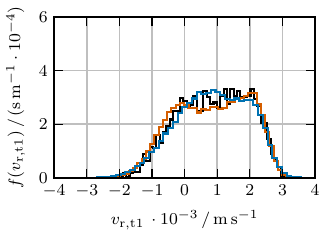}
            %\inputtikz{figures}{reflected_velocity_distribution_t1_3000.0_60}
        \end{minipage}\hfill%
        \begin{minipage}[b]{0.32\textwidth}
            \centering
            \includegraphics{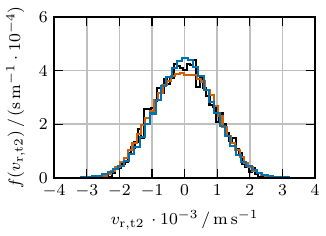}
            %\inputtikz{figures}{reflected_velocity_distribution_t2_3000.0_60}
        \end{minipage}\hfill%
        \begin{minipage}[b]{0.32\textwidth}
            \centering
            \includegraphics{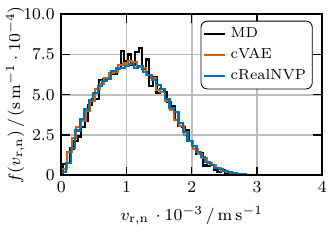}
            %\inputtikz{figures}{reflected_velocity_distribution_n_3000.0_60}
        \end{minipage}
        \vspace{-2mm}
        \subcaption{$|\bm{v}_{\text{i}}| = 3000.0 \, \text{m}/\text{s}$, $\theta_{\text{i}} = 60.0 \, ^\circ$}
    \end{minipage}
    \begin{minipage}{\textwidth}
        \centering
        \begin{minipage}[b]{0.32\textwidth}
            \centering
            \includegraphics{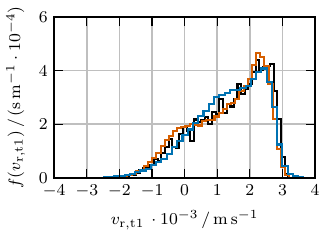}
            %\inputtikz{figures}{reflected_velocity_distribution_t1_3000.0_80}
        \end{minipage}\hfill%
        \begin{minipage}[b]{0.32\textwidth}
            \centering
            \includegraphics{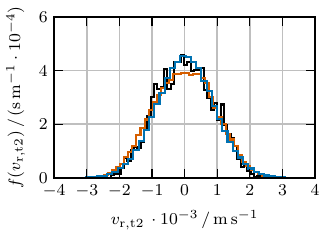}
            %\inputtikz{figures}{reflected_velocity_distribution_t2_3000.0_80}
        \end{minipage}\hfill%
        \begin{minipage}[b]{0.32\textwidth}
            \centering
            \includegraphics{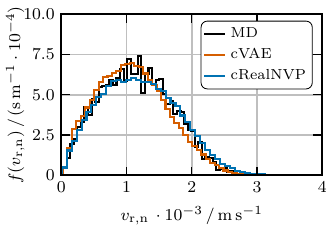}
            %\inputtikz{figures}{reflected_velocity_distribution_n_3000.0_80}
        \end{minipage}
        \vspace{-2mm}
        \subcaption{$|\bm{v}_{\text{i}}| = 3000.0 \, \text{m}/\text{s}$, $\theta_{\text{i}} = 80.0 \, ^\circ$}
    \end{minipage}
    \vspace{-2mm}
    \caption{Reflected velocity distributions of atomic oxygen for the \gls{cvae}, \gls{crealnvp} and \gls{md} data for an incident velocity magnitude of $|\vecvi| = 3000.0\,\text{m/s}$ and varying polar angles $\theta_{\text{i}}$ at a wall temperature of $\Tw = 300\,\text{K}$.}
    \label{fig:histograms_v_mag_3000}
\end{figure}

\newpage

\begin{figure}[H]
    \centering
    \begin{minipage}{\textwidth}
        \centering
        \begin{minipage}[b]{0.32\textwidth}
            \centering
            \includegraphics{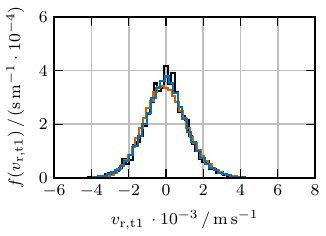}
            %\inputtikz{figures}{reflected_velocity_distribution_t1_7261.3_0}
        \end{minipage}\hfill%
        \begin{minipage}[b]{0.32\textwidth}
            \centering
            \includegraphics{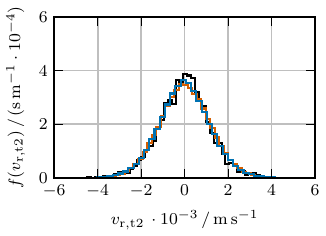}
            %\inputtikz{figures}{reflected_velocity_distribution_t2_7261.3_0}
        \end{minipage}\hfill%
        \begin{minipage}[b]{0.32\textwidth}
            \centering
            \includegraphics{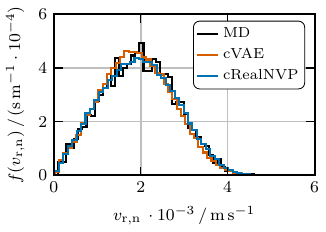}
            %\inputtikz{figures}{reflected_velocity_distribution_n_7261.3_0}
        \end{minipage}
        \vspace{-2mm}
        \subcaption{$|\bm{v}_{\text{i}}| = 7261.3 \, \text{m}/\text{s}$, $\theta_{\text{i}} = 0.0 \, ^\circ$}
    \end{minipage}
    \begin{minipage}{\textwidth}
        \centering
        \begin{minipage}[b]{0.32\textwidth}
            \centering
            \includegraphics{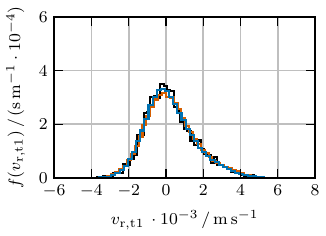}
            %\inputtikz{figures}{reflected_velocity_distribution_t1_7261.3_20}
        \end{minipage}\hfill%
        \begin{minipage}[b]{0.32\textwidth}
            \centering
            \includegraphics{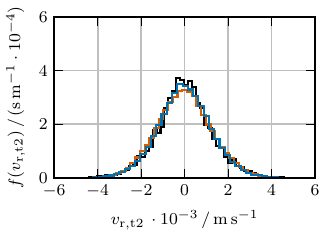}
            %\inputtikz{figures}{reflected_velocity_distribution_t2_7261.3_20}
        \end{minipage}\hfill%
        \begin{minipage}[b]{0.32\textwidth}
            \centering
            \includegraphics{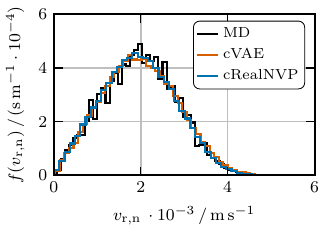}
            %\inputtikz{figures}{reflected_velocity_distribution_n_7261.3_20}
        \end{minipage}
        \vspace{-2mm}
        \subcaption{$|\bm{v}_{\text{i}}| = 7261.3 \, \text{m}/\text{s}$, $\theta_{\text{i}} = 20.0 \, ^\circ$}
    \end{minipage}
    \begin{minipage}{\textwidth}
        \centering
        \begin{minipage}[b]{0.32\textwidth}
            \centering
            \includegraphics{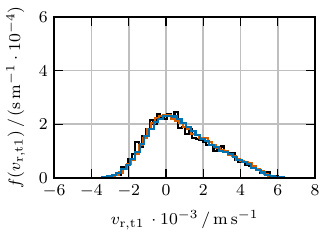}
            %\inputtikz{figures}{reflected_velocity_distribution_t1_7261.3_40}
        \end{minipage}\hfill%
        \begin{minipage}[b]{0.32\textwidth}
            \centering
            \includegraphics{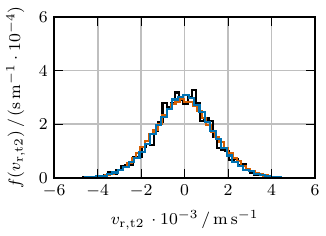}
            %\inputtikz{figures}{reflected_velocity_distribution_t2_7261.3_40}
        \end{minipage}\hfill%
        \begin{minipage}[b]{0.32\textwidth}
            \centering
            \includegraphics{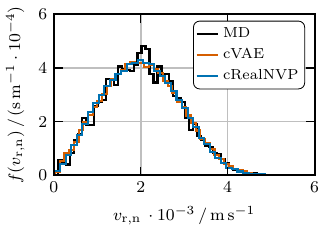}
            %\inputtikz{figures}{reflected_velocity_distribution_n_7261.3_40}
        \end{minipage}
        \vspace{-2mm}
        \subcaption{$|\bm{v}_{\text{i}}| = 7261.3 \, \text{m}/\text{s}$, $\theta_{\text{i}} = 40.0 \, ^\circ$}
    \end{minipage}
    \begin{minipage}{\textwidth}
        \centering
        \begin{minipage}[b]{0.32\textwidth}
            \centering
            \includegraphics{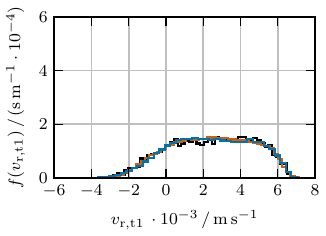}
            %\inputtikz{figures}{reflected_velocity_distribution_t1_7261.3_60}
        \end{minipage}\hfill%
        \begin{minipage}[b]{0.32\textwidth}
            \centering
            \includegraphics{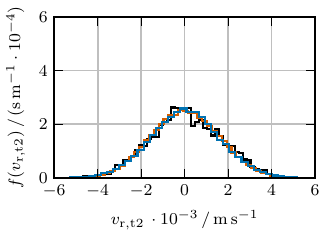}
            %\inputtikz{figures}{reflected_velocity_distribution_t2_7261.3_60}
        \end{minipage}\hfill%
        \begin{minipage}[b]{0.32\textwidth}
            \centering
            \includegraphics{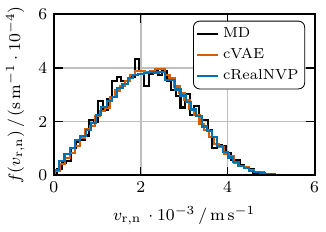}
            %\inputtikz{figures}{reflected_velocity_distribution_n_7261.3_60}
        \end{minipage}
        \vspace{-2mm}
        \subcaption{$|\bm{v}_{\text{i}}| = 7261.3 \, \text{m}/\text{s}$, $\theta_{\text{i}} = 60.0 \, ^\circ$}
    \end{minipage}
    \begin{minipage}{\textwidth}
        \centering
        \begin{minipage}[b]{0.32\textwidth}
            \centering
            \includegraphics{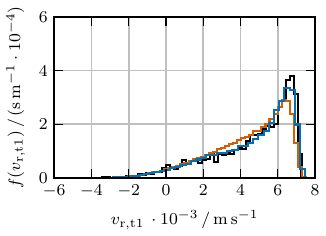}
            %\inputtikz{figures}{reflected_velocity_distribution_t1_7261.3_80}
        \end{minipage}\hfill%
        \begin{minipage}[b]{0.32\textwidth}
            \centering
            \includegraphics{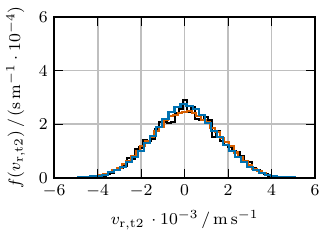}
            %\inputtikz{figures}{reflected_velocity_distribution_t2_7261.3_80}
        \end{minipage}\hfill%
        \begin{minipage}[b]{0.32\textwidth}
            \centering
            \includegraphics{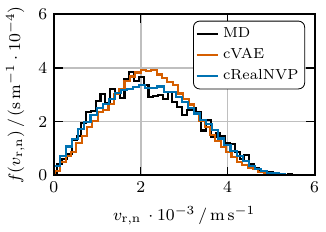}
            %\inputtikz{figures}{reflected_velocity_distribution_n_7261.3_80}
        \end{minipage}
        \vspace{-2mm}
        \subcaption{$|\bm{v}_{\text{i}}| = 7261.3 \, \text{m}/\text{s}$, $\theta_{\text{i}} = 80.0 \, ^\circ$}
        \label{subfig:histogram_v_mag_7261.3_80}
    \end{minipage}
    \vspace{-2mm}
    \caption{Reflected velocity distributions of atomic oxygen for the \gls{cvae}, \gls{crealnvp} and \gls{md} data for an incident velocity magnitude of $|\vecvi| = 7261.3\,\text{m/s}$ and varying polar angles $\theta_{\text{i}}$ at a wall temperature of $\Tw = 300\,\text{K}$.}
    \label{fig:histograms_v_mag_7261.3}
\end{figure}
\newpage

\begin{figure}[H]
    \centering
    \begin{minipage}{\textwidth}
        \centering
        \begin{minipage}[b]{0.32\textwidth}
            \centering
            \includegraphics{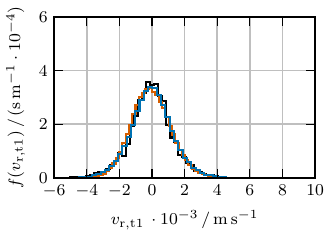}
            %\inputtikz{figures}{reflected_velocity_distribution_t1_8585.9_0}
        \end{minipage}\hfill%
        \begin{minipage}[b]{0.32\textwidth}
            \centering
            \includegraphics{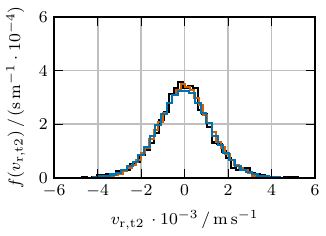}
            %\inputtikz{figures}{reflected_velocity_distribution_t2_8585.9_0}
        \end{minipage}\hfill%
        \begin{minipage}[b]{0.32\textwidth}
            \centering
            \includegraphics{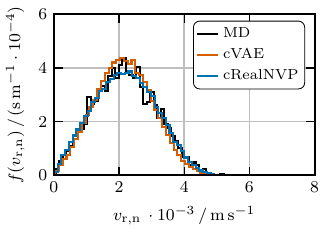}
            %\inputtikz{figures}{reflected_velocity_distribution_n_8585.9_0}
        \end{minipage}
        \vspace{-2mm}
        \subcaption{$|\bm{v}_{\text{i}}| = 8585.9 \, \text{m}/\text{s}$, $\theta_{\text{i}} = 0.0 \, ^\circ$}
    \end{minipage}
    \begin{minipage}{\textwidth}
        \centering
        \begin{minipage}[b]{0.32\textwidth}
            \centering
            \includegraphics{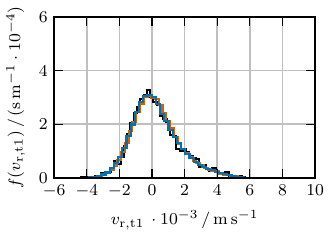}
            %\inputtikz{figures}{reflected_velocity_distribution_t1_8585.9_20}
        \end{minipage}\hfill%
        \begin{minipage}[b]{0.32\textwidth}
            \centering
            \includegraphics{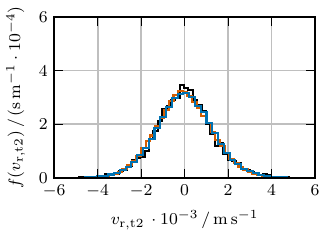}
            %\inputtikz{figures}{reflected_velocity_distribution_t2_8585.9_20}
        \end{minipage}\hfill%
        \begin{minipage}[b]{0.32\textwidth}
            \centering
            \includegraphics{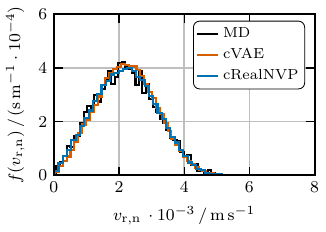}
            %\inputtikz{figures}{reflected_velocity_distribution_n_8585.9_20}
        \end{minipage}
        \vspace{-2mm}
        \subcaption{$|\bm{v}_{\text{i}}| = 8585.9 \, \text{m}/\text{s}$, $\theta_{\text{i}} = 20.0 \, ^\circ$}
    \end{minipage}
    \begin{minipage}{\textwidth}
        \centering
        \begin{minipage}[b]{0.32\textwidth}
            \centering
            \includegraphics{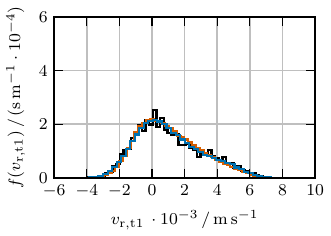}
            %\inputtikz{figures}{reflected_velocity_distribution_t1_8585.9_40}
        \end{minipage}\hfill%
        \begin{minipage}[b]{0.32\textwidth}
            \centering
            \includegraphics{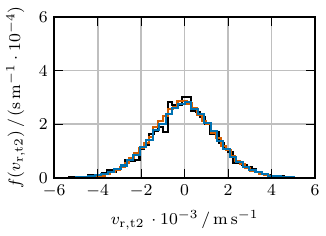}
            %\inputtikz{figures}{reflected_velocity_distribution_t2_8585.9_40}
        \end{minipage}\hfill%
        \begin{minipage}[b]{0.32\textwidth}
            \centering
            \includegraphics{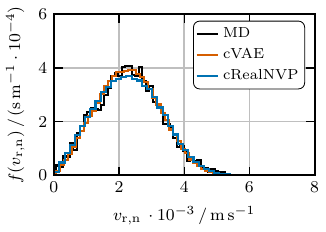}
            %\inputtikz{figures}{reflected_velocity_distribution_n_8585.9_40}
        \end{minipage}
        \vspace{-2mm}
        \subcaption{$|\bm{v}_{\text{i}}| = 8585.9 \, \text{m}/\text{s}$, $\theta_{\text{i}} = 40.0 \, ^\circ$}
    \end{minipage}
    \begin{minipage}{\textwidth}
        \centering
        \begin{minipage}[b]{0.32\textwidth}
            \centering
            \includegraphics{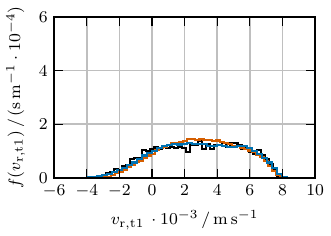}
            %\inputtikz{figures}{reflected_velocity_distribution_t1_8585.9_60}
        \end{minipage}\hfill%
        \begin{minipage}[b]{0.32\textwidth}
            \centering
            \includegraphics{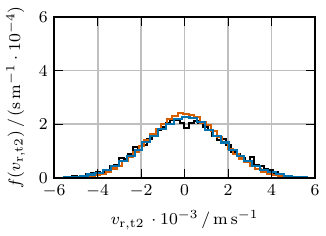}
            %\inputtikz{figures}{reflected_velocity_distribution_t2_8585.9_60}
        \end{minipage}\hfill%
        \begin{minipage}[b]{0.32\textwidth}
            \centering
            \includegraphics{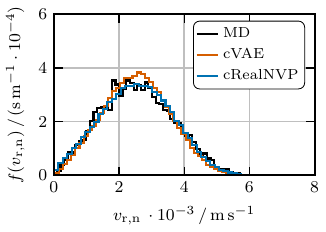}
            %\inputtikz{figures}{reflected_velocity_distribution_n_8585.9_60}
        \end{minipage}
        \vspace{-2mm}
        \subcaption{$|\bm{v}_{\text{i}}| = 8585.9 \, \text{m}/\text{s}$, $\theta_{\text{i}} = 60.0 \, ^\circ$}
    \end{minipage}
    \begin{minipage}{\textwidth}
        \centering
        \begin{minipage}[b]{0.32\textwidth}
            \centering
            \includegraphics{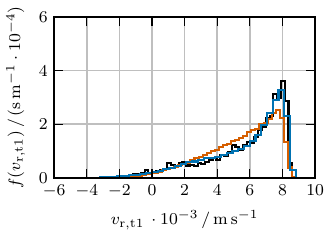}
            %\inputtikz{figures}{reflected_velocity_distribution_t1_8585.9_80}
        \end{minipage}\hfill%
        \begin{minipage}[b]{0.32\textwidth}
            \centering
            \includegraphics{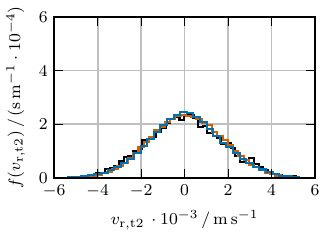}
            %\inputtikz{figures}{reflected_velocity_distribution_t2_8585.9_80}
        \end{minipage}\hfill%
        \begin{minipage}[b]{0.32\textwidth}
            \centering
            \includegraphics{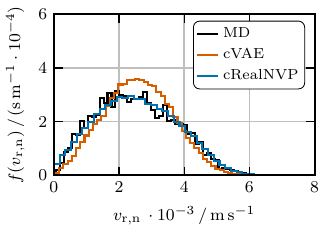}
            %\inputtikz{figures}{reflected_velocity_distribution_n_8585.9_80}
        \end{minipage}
        \vspace{-2mm}
        \subcaption{$|\bm{v}_{\text{i}}| = 8585.9 \, \text{m}/\text{s}$, $\theta_{\text{i}} = 80.0 \, ^\circ$}
        \label{subfig:histogram_v_mag_8585.9_80}
    \end{minipage}
    \vspace{-2mm}
    \caption{Reflected velocity distributions of atomic oxygen for the \gls{cvae}, \gls{crealnvp} and \gls{md} data for an incident velocity magnitude of $|\vecvi| = 8585.9\,\text{m/s}$ and varying polar angles $\theta_{\text{i}}$ at a wall temperature of $\Tw = 300\,\text{K}$.}
    \label{fig:histograms_v_mag_8585.9}
\end{figure}

\newpage

\section{Non-Equilibrium Validation Across Wall Temperatures} \label{secapdix:neq_validation_wall_temp}
\begin{figure}[H]
    \centering
    \begin{minipage}{\textwidth}
        \centering
        \begin{minipage}[b]{0.32\textwidth}
            \centering
            \includegraphics{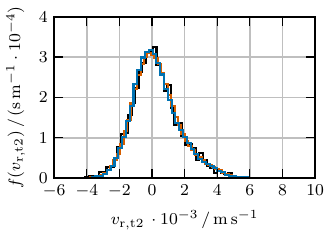}
            %\inputtikz{figures}{reflected_velocity_distribution_t1_7800.0_20_900}
        \end{minipage}\hfill%
        \begin{minipage}[b]{0.32\textwidth}
            \centering
            \includegraphics{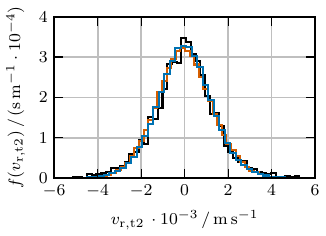}
            %\inputtikz{figures}{reflected_velocity_distribution_t2_7800.0_20_900}
        \end{minipage}\hfill%
        \begin{minipage}[b]{0.32\textwidth}
            \centering
            \includegraphics{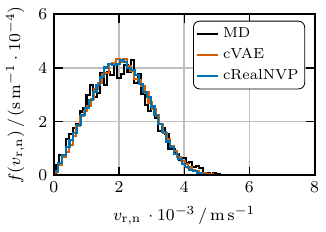}
            %\inputtikz{figures}{reflected_velocity_distribution_n_7800.0_20_900}
        \end{minipage}
        \vspace{-2mm}
        \subcaption{$|\bm{v}_{\text{i}}| = 7800.0 \, \text{m}/\text{s}$, $\theta_{\text{i}} = 20.0 \, ^\circ$, $\Tw = 900\,\text{K}$}
    \end{minipage}
    \begin{minipage}{\textwidth}
        \centering
        \begin{minipage}[b]{0.32\textwidth}
            \centering
            \includegraphics{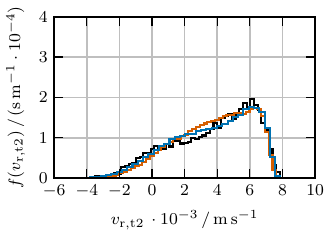}
            %\inputtikz{figures}{reflected_velocity_distribution_t1_7800.0_70_900}
        \end{minipage}\hfill%
        \begin{minipage}[b]{0.32\textwidth}
            \centering
            \includegraphics{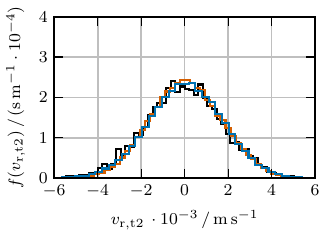}
            %\inputtikz{figures}{reflected_velocity_distribution_t2_7800.0_70_900}
        \end{minipage}\hfill%
        \begin{minipage}[b]{0.32\textwidth}
            \centering
            \includegraphics{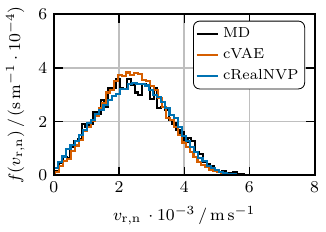}
            %\inputtikz{figures}{reflected_velocity_distribution_n_7800.0_70_900}
        \end{minipage}
        \vspace{-2mm}
        \subcaption{$|\bm{v}_{\text{i}}| = 7800.0 \, \text{m}/\text{s}$, $\theta_{\text{i}} = 70.0 \, ^\circ$, $\Tw = 900\,\text{K}$}
    \end{minipage}
    \begin{minipage}{\textwidth}
        \centering
        \begin{minipage}[b]{0.32\textwidth}
            \centering
            \includegraphics{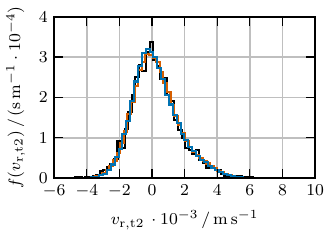}
            %\inputtikz{figures}{reflected_velocity_distribution_t1_7800.0_20_1500}
        \end{minipage}\hfill%
        \begin{minipage}[b]{0.32\textwidth}
            \centering
            \includegraphics{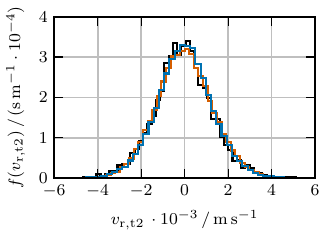}
            %\inputtikz{figures}{reflected_velocity_distribution_t2_7800.0_20_1500}
        \end{minipage}\hfill%
        \begin{minipage}[b]{0.32\textwidth}
            \centering
            \includegraphics{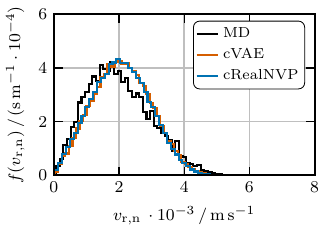}
            %\inputtikz{figures}{reflected_velocity_distribution_n_7800.0_20_1500}
        \end{minipage}
        \vspace{-2mm}
        \subcaption{$|\bm{v}_{\text{i}}| = 7800.0 \, \text{m}/\text{s}$, $\theta_{\text{i}} = 20.0 \, ^\circ$, $\Tw = 1\,500\,\text{K}$}
    \end{minipage}
    \begin{minipage}{\textwidth}
        \centering
        \begin{minipage}[b]{0.32\textwidth}
            \centering
            \includegraphics{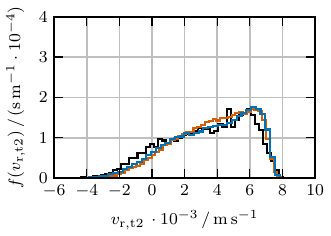}
            %\inputtikz{figures}{reflected_velocity_distribution_t1_7800.0_70_1500}
        \end{minipage}\hfill%
        \begin{minipage}[b]{0.32\textwidth}
            \centering
            \includegraphics{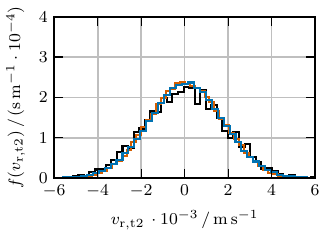}
            %\inputtikz{figures}{reflected_velocity_distribution_t2_7800.0_70_1500}
        \end{minipage}\hfill%
        \begin{minipage}[b]{0.32\textwidth}
            \centering
            \includegraphics{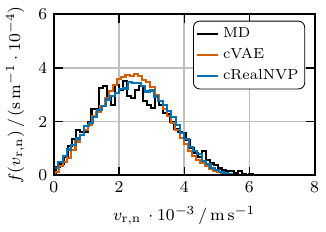}
            %\inputtikz{figures}{reflected_velocity_distribution_n_7800.0_70_1500}
        \end{minipage}
        \vspace{-2mm}
        \subcaption{$|\bm{v}_{\text{i}}| = 7800.0 \, \text{m}/\text{s}$, $\theta_{\text{i}} = 70.0 \, ^\circ$, $\Tw = 1\,500\,\text{K}$}
    \end{minipage}
    \vspace{-2mm}
    \caption{Reflected velocity distributions of atomic oxygen for the \gls{cvae}, \gls{crealnvp} and \gls{md} data for an incident velocity magnitude of $|\vecvi| = 7800.0\,\text{m/s}$, varying polar angles $\theta_{\text{i}}$, and two different wall temperatures $\Tw$.}
    \label{fig:histograms_v_mag_7800.0_different_Tw}
\end{figure}

\newpage

\section{Reservoir Application Across Wall Temperatures and Species} \label{secapdix:reservoir_application}

\begin{figure}[H]
    \centering
    \begin{minipage}[b]{0.45\textwidth}
        \centering
        \includegraphics{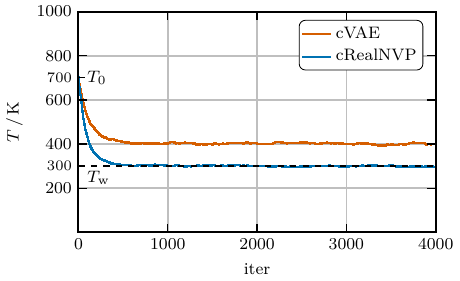}
        %\inputtikz{figures}{reservoir_gas_temperature_O2_700_300}
        \subcaption{Molecular oxygen and $\Tw = 300\,\text{K}$}
    \end{minipage}\hfill%
    \begin{minipage}[b]{0.45\textwidth}
        \centering
        \includegraphics{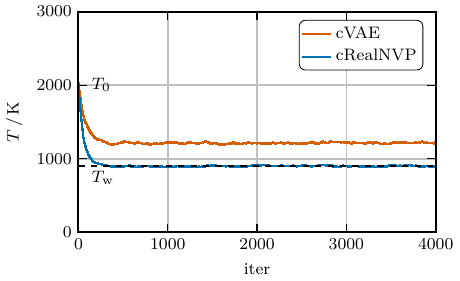}
        %\inputtikz{figures}{reservoir_gas_temperature_O2_2000_900}
        \subcaption{Molecular oxygen and $\Tw = 900\,\text{K}$}
    \end{minipage}
    \\[6pt]
    \begin{minipage}[b]{0.45\textwidth}
        \centering
        \includegraphics{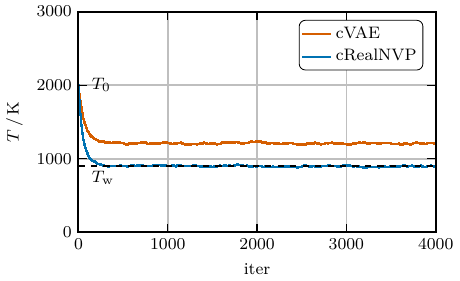}
        %\inputtikz{figures}{reservoir_gas_temperature_N2_2000_900}
        \subcaption{Molecular nitrogen and $\Tw = 900\,\text{K}$}
    \end{minipage}\hfill%
    \begin{minipage}[b]{0.45\textwidth}
        \centering
        \includegraphics{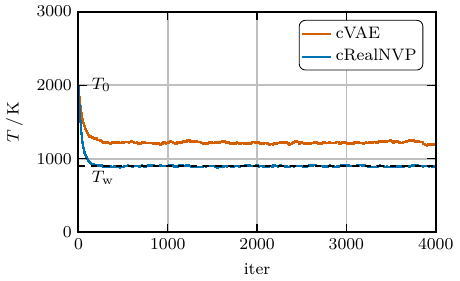}
        %\inputtikz{figures}{reservoir_gas_temperature_O_2000_900}
        \subcaption{Atomic oxygen and $\Tw = 900\,\text{K}$}
    \end{minipage}
    \\[6pt]
    \begin{minipage}[b]{0.45\textwidth}
        \centering
        \includegraphics{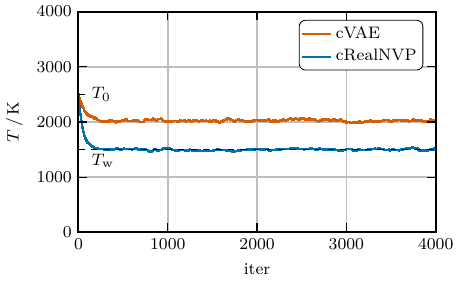}
        %\inputtikz{figures}{reservoir_gas_temperature_O2_2500_1500}
        \subcaption{Molecular oxygen and $\Tw = 1500\,\text{K}$}
    \end{minipage}\hfill%
    \begin{minipage}[b]{0.45\textwidth}
        \centering
        \includegraphics{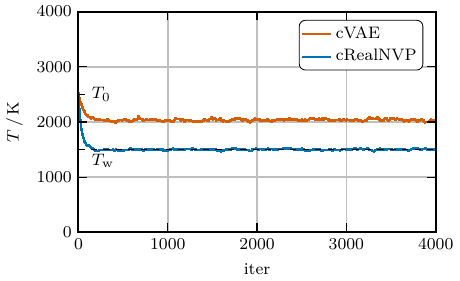}
        %\inputtikz{figures}{reservoir_gas_temperature_O_2500_1500}
        \subcaption{Atomic oxygen and $\Tw = 1500\,\text{K}$}
    \end{minipage}
    \caption{Evolution of the gas temperature in the reservoir simulation for different species, initial temperatures $T_0$ and wall temperatures $T_{\text{w}}$ for the \gls{cvae} and the \gls{crealnvp} models.}
    \label{fig:reservoir_gas_temperature_appx}
\end{figure}

\end{document}

%% file: acronyms.tex
\newacronym{md}{MD}{molecular dynamics}
\newacronym{gsi}{GSI}{gas-surface interaction}
\newacronym{dsmc}{DSMC}{direct simulation Monte Carlo}
\newacronym{cvae}{cVAE}{conditional Variational Autoencoder}
\newacronym{uq}{UQ}{uncertainty quantification}
\newacronym{ao}{AO}{atomic oxygen}
\newacronym{mfmc}{MFMC}{multi-fidelity Monte Carlo}
\newacronym{crealnvp}{cRealNVP}{conditional Real-valued Non-Volume Preserving}
\newacronym{db}{DB}{detailed balance}
\newacronym{fwd}{fwd}{forward}
\newacronym{bwd}{bwd}{backward}
\newacronym{realnvp}{RealNVP}{Real-valued Non-Volume Preserving}
\newacronym{nll}{NLL}{negative log-likelihood}
\newacronym{kl}{KL}{Kulback-Leibler}
\newacronym{abep}{ABEP}{air-breathing electric propulsion}
\newacronym{cl}{CL}{Cercignani-Lampis}
\newacronym{imd}{IMD}{ITAP Molecular Dynamics}
\newacronym{dft}{DFT}{density functional theory}
\newacronym{tpmc}{TPMC}{test particle Monte Carlo}
\newacronym{elu}{ELU}{exponential linear unit}
\newacronym{relu}{ReLU}{rectified linear unit}
\newacronym{leakyrelu}{leaky ReLU}{leaky rectified linear unit}
\newacronym{mse}{MSE}{mean squared error}
\newacronym{tanh}{tanh}{hyperbolic tangent}

\newglossaryentry{leo}
{
  name={LEO},
  description={low Earth orbit},
  first={low Earth orbit (LEO)},
  text={LEO}
}
\newglossaryentry{vleo}
{
  name={VLEO},
  description={very low Earth orbit},
  first={very low Earth orbit (VLEO)},
  text={VLEO}
}